\PassOptionsToPackage{pdfpagelabels=false}{hyperref}
\documentclass[fleqn,usenatbib]{mnras}

\usepackage[T1]{fontenc}
\usepackage{ae,aecompl}

\usepackage{float}
\usepackage{hyperref}
\usepackage{apjfonts}
\usepackage{amssymb}
\usepackage{amsmath}
\usepackage{graphicx}	
\usepackage[normalem]{ulem}
\usepackage{multirow}
\usepackage{mathtools}
\usepackage{dutchcal}
\usepackage{supertabular}
\usepackage[all]{hypcap}
\hypersetup{colorlinks,citecolor=blue}

\usepackage{etoolbox}
\makeatletter
\patchcmd\@combinedblfloats{\box\@outputbox}{\unvbox\@outputbox}{}{%
  \errmessage{\noexpand\@combinedblfloats could not be patched}%
}%
\makeatother

\newcommand{\mystar}{\ast}        
\newcommand\run[1]{\texttt{#1}}   
\newcommand{\sigstar}{\sigma_\mystar}
\newcommand{\mvir}{M_\mathrm{vir}}

\newcommand{\rcontam}{r_\mathrm{contam}}

\newcommand{\vmax}{V_\mathrm{max}}

\newcommand{\msun}{\textnormal{M}_\odot}

\newcommand{\mstar}{M_\mystar}

\newcommand{\mhalo}{M_\mathrm{halo}}

\newcommand{\mpc}{\mathrm{Mpc}}

\newcommand{\kpc}{\mathrm{kpc}}
\newcommand{\pc}{\mathrm{pc}}

\newcommand{\hmpc}{h^{-1}\,\mathrm{Mpc}}

\newcommand{\kms}{{\rm km} \ {\rm s}^{-1}}

\newcommand{\vcirc}{V_\mathrm{circ}}
\newcommand{\rmax}{R_\mathrm{max}}

\newcommand{\vhalf}{V_\mathrm{1/2}}
\newcommand{\rhalf}{R_\mathrm{1/2}}

\newcommand{\raj}{\run{Romeo} \run{\&} \run{Juliet}}
\newcommand{\tal}{\run{Thelma} \run{\&} \run{Louise}}

\newcommand{\egas}{\epsilon^\mathrm{gas}_\mathrm{min}}

\newcommand{\etal}{et al.}

\newcommand\altaffilmark[1]{$^{#1}$}
\newcommand\altaffiltext[1]{$^{#1}$}

\title[The Local Group on FIRE]{The Local Group on FIRE:  Dwarf galaxy populations across a suite of hydrodynamic simulations
\vspace{-0.5cm}}

\vspace{-0.2cm}
\author[S. Garrison-Kimmel \etal]{
\parbox[t]{\textwidth}{
Shea Garrison-Kimmel\thanks{$\!$sheagk@caltech.edu}\thanks{Einstein Fellow}\altaffilmark{1},
Philip F.~Hopkins\altaffilmark{1},
Andrew Wetzel\thanks{Caltech-Carnegie Fellow}\altaffilmark{1,2,3},
James S.~Bullock\altaffilmark{4},
Michael Boylan-Kolchin\altaffilmark{5},
Du{\v s}an Kere{\v s}\altaffilmark{6},
Claude-Andr{\'e} Faucher-Gigu{\`e}re\altaffilmark{7},
Kareem El-Badry\altaffilmark{8},
Astrid Lamberts\altaffilmark{1},
Eliot Quataert\altaffilmark{8},
Robyn Sanderson\altaffilmark{1}
}
\vspace*{6pt} \\
\altaffiltext{1}{TAPIR, Mailcode 350-17, California Institute of Technology, Pasadena, CA 91125, USA} \\
\altaffiltext{2}{The Observatories of the Carnegie Institution for Science, Pasadena, CA 91101, USA} \\
\altaffiltext{3}{Department of Physics, University of California, Davis, CA 95616, USA} \\
\altaffiltext{4}{Center for Cosmology, Department of Physics and Astronomy, University of California, Irvine, CA 92697, USA} \\
\altaffiltext{5}{Department of Astronomy, The University of Texas at Austin, 2515 Speedway, Stop C1400, Austin, TX 78712, USA} \\
\altaffiltext{6}{Department of Physics, Center for Astrophysics and Space Science, University of California at San Diego, 9500 Gilman Drive, La Jolla, CA 92093} \\
\altaffiltext{7}{Department of Physics and Astronomy and CIERA, Northwestern University, 2145 Sheridan Road, Evanston, IL 60208, USA} \\
\altaffiltext{8}{Department of Astronomy and Theoretical Astrophysics Center, University of California Berkeley, Berkeley, CA 94720} \\
\vspace{-0.5cm}
}

\date{Accepted XXX. Received YYY; in original form ZZZ}

\pubyear{2018}

\begin{document}
\label{firstpage}
\pagerange{\pageref{firstpage}--\pageref{lastpage}}

\maketitle 
\begin{abstract}

We present a new set of high-resolution hydrodynamic cosmological zoom-in 
simulations that apply the Feedback In Realistic Environments (FIRE) physics 
to both Local Group (LG)-like and isolated Milky Way (MW)-like volumes 
(ten host systems in total with baryonic particle mass $\simeq 3,500-7,000\,\msun$).
We study the stellar mass functions, circular velocity or mass profiles, and
velocity dispersions of the dwarf galaxy populations. The simulations
reproduce the stellar mass function and central densities of MW satellite dwarfs
for $\mstar\geq10^{5.5}\,\msun$ and predict the existence of $\sim3$
unidentified galaxies with $\mstar\sim10^5\,\msun$ within $300~\kpc$ of the MW.
Overall, we find no evidence for the classical missing satellites or too-big-to-fail 
(TBTF) problems for satellite galaxies in our sample. Among the satellites, 
TBTF is resolved primarily by subhalo disruption and overall mass loss; central 
density profiles of subhalos are of secondary importance.  For non-satellite 
galaxies, our LG-like simulations predict as many as $\sim10$ as-of-yet unseen 
galaxies at distances $0.3-1\,\mpc$ from both hosts, with 
$\mstar\simeq10^{5-6}\,\msun$ (in halos with $\vmax \sim 20~\kms$), albeit with 
large halo-to-halo variance.  None of our simulations produces a compact, 
baryon-dominated, high-density dwarf elliptical-type galaxy (with $\vcirc \gtrsim 35~\kms$ 
at $r<1\,$kpc), of which six may appear in the LG (but none in the MW). It 
may therefore remain a challenge to reproduce the full diversity of the dwarf 
population, including both the highest and lowest density systems.

\end{abstract}




\section{Introduction}
\label{sec:intro}
Our location within the Local Group (LG) affords it a unique importance
in astronomy.  It remains the only part of the Universe where we can detect
tiny dwarf galaxies (stellar mass $\mstar\lesssim10^6\msun$), let alone use
resolved stellar observations to study their internal properties and
kinematics.  As the most dark matter-dominated galaxies in the Universe
\citep[e.g.][]{McConnachie2012}, these dwarf galaxies provide crucial tests
of the standard structure formation paradigm, cold dark matter with a cosmological
constant ($\Lambda$CDM), and may ultimately indirectly reveal the nature of DM
itself \citep[e.g.][]{fermidwarfs}.

While $\Lambda$CDM reproduces large-scale observations extraordinarily
well \citep[e.g.][]{millennium}, explaining the dwarf galaxy population
within the $\Lambda$CDM framework has historically proven difficult (see
\citealp{BBK2017} for a recent review).  Perhaps most famously, the ``missing
satellites'' problem (MSP; \citealp{Moore1999,Klypin1999MissingSats}) points out
that dark matter-only (DMO) simulations of MW-mass hosts in $\Lambda$CDM
predict orders of magnitude more bound subhalos within $\sim300~\kpc$ than
known luminous satellites of the MW.  While the MSP is
usually accounted for by a combination of photoionization during reionization
\citep{Bullock2000,Somerville2002}, observational bias and incompleteness
\citep[e.g.][]{Tollerud2008}, and subhalo destruction due to the MW disk
\citep[][]{DOnghia2010,Sawala2016b,GKDisk}, these solutions typically
resolve the disparity by placing the known MW satellites in the largest
subhalos predicted around MW-mass hosts and leaving the smallest clumps
undetected or entirely dark.  This picture is further supported by the
success of applying extrapolations of the abundance matching paradigm,
which successfully reproduces large-scale clustering statistics by assuming
a relatively tight relationship between halo mass $\mhalo$ and $\mstar$,
to the LG environment \citep{ELVIS}.  

However, the too-big-to-fail (TBTF) problem notes that the circular velocity
profiles of the largest subhalos in DMO simulations of MW-mass galaxies (i.e.
the subhalos assumed to host the luminous satellites) are incompatible with
observational constraints on the MW dwarf satellites \citep{BK2011,BK2012}.  A
similar discrepancy exists when comparing with the satellite galaxies of M31
\citep{Tollerud2014} or the dwarf galaxies in the Local Field (defined here as
within $1~\mpc$ of the MW or M31, but more than $300~\kpc$ from both;
\citealp{ELVISTBTF}), and TBTF even appears to exist beyond the LG entirely
\citep{Papastergis2015,Papastergis2016}:  dwarf galaxies
($\mstar\sim10^{5-7}\msun$) have less mass within $\sim250~\pc-1~\kpc$ than
DMO simulations of the halos expected to host those galaxies predict.

Recent simulations have begun to jointly resolve the MSP\footnote{The Auriga
simulations \citep{Grand2017}, high-resolution magneto-hydrodynamic zoom-ins
focusing on isolated MW-mass galaxies, also reproduce the MW/M31 satellite
luminosity functions down to $5\times10^5\msun$ \citep{Simpson2017}, though to
date there have been no analyses of the internal structure of those satellites.}
and TBTF by more realistically modeling gas cooling, star formation, and
stellar/supernovae feedback.  For example, \citet{BrooksZolotov2012}, using
simulations from \citet{Zolotov2012}, demonstrated a reduction in the peak
circular velocity of the halos associated with TBTF due to a combination of
supernovae feedback \citep[modeled via the ``blastwave'' scheme of][]{Stinson2006} 
and tidal disruption, such that their simulations were free of both TBTF and 
the MSP.  More recently, \citet{Dutton2016} and \citet{Buck2018} showed that 
the NIHAO simulation suite, which also adopts the blastwave scheme, is similarly 
free of the MSP and TBTF.

The conclusion that TBTF and the MSP can be explained via baryonic physics,
even using non-blastwave feedback implementations, is growing increasingly
robust.  The APOSTLE simulations \citep{Fattahi2016,Sawala2016}, for example,
apply the EAGLE models for galaxy formation, which are tuned to reproduce the
stellar mass function and sizes of galaxies at $z=0.1$
\citep{Crain2015,Schaye2015}, to 12 LG-like volumes,\footnote{The APOSTLE
simulations follow in the spiritual footsteps of the CLUES (Constrained Local
UniversE Simulations) project \citep[e.g.][]{Gottloeber2010} in targeting 
LG-like pairs in hydrodynamic, cosmological zoom-in simulations.  The CLUES
simulations, however, constrain the $\sim5~\hmpc$ environment around the
targeted hosts to match that of the actual  LG.} demonstrating that
extrapolations of models that match the statistics of larger galaxies can also
duplicate the LG.  The APOSTLE dwarf galaxy populations generally do not exhibit
the MSP: the simulated volumes contain  a similar number of galaxies with
$\mstar\geq10^5\msun$ as the actual MW, M31, and LG. Moreover, the mass function
of subhalos that host the luminous dwarf galaxies in APOSTLE (quantified by
$\vmax$, the peak of the circular velocity curve) agree with the mass function
implied by the \citet{Penarrubia2008} estimates for the MW dwarf spheroidals
(dSphs), implying that the APOSTLE hosts are also free of the TBTF problem.

In an alternative approach, \citet{Wetzel2016} used the Feedback In Realistic
Environments
\citep[FIRE;][]{FIRE,FIRE2}\footnote{\url{http://fire.northwestern.edu}} physics
to simulate an isolated MW-mass galaxy with high enough resolution to capture
the internal dynamics of the classical satellites.  FIRE includes explicit
models for star formation and stellar/supernovae feedback that self-consistently
yield bursty star formation in dwarf galaxies \citep{Muratov2015,
Sparre2017,CAFG2017,ElBadry2016} and overall agreement with a variety of 
galaxy-scale observables, including the star formation histories of dwarf galaxies
\citep{Onorbe2015,Wetzel2016,Fitts2017}; the mass--metallicity \citep{Ma2016a}, 
stellar mass--halo mass \citep{FIRE,FIRE2}, and stellar mass--star formation 
rate \citep{Sparre2017} relationships; and the fraction of the stellar mass 
in the halos of MW-mass galaxies \citep{Sanderson2017}. \citet{Wetzel2016} showed 
that FIRE also yields a reasonable MW satellite population:  the set of simulated 
dwarf galaxies falls roughly midway between that of the MW and M31 when counting 
galaxies either by $\mstar$ or by the line-of-sight stellar velocity dispersion 
$\sigstar$, the observable relevant to TBTF.

These works, however, have suffered from limitations.  While the hosts in the 
APOSTLE simulations are carefully selected to match the LG environment, 
the majority of the APOSTLE results are drawn from their `L2' simulations with
baryonic particle masses $\sim10^5\msun$, approaching the total mass of the
smaller classical dwarf galaxies.  In addition, the effective equation of state
and the spatial/density resolution used in the APOSTLE simulations is such that
the smallest resolvable Jeans/Toomre mass is $>10^8\msun$; therefore, clouds in
lower mass galaxies cannot be self-consistently resolved.  The simulations in
\citet{Zolotov2012} and \citet{Buck2018} similarly have baryonic particle masses
$>20,000~\msun$, with the highest resolutions reached at lower halo masses
$\sim8\times10^{11}\msun$. \citet{Wetzel2016} reached higher resolutions and
used a more physical subgrid model for star formation and feedback, but their
results are based on a single simulation of an isolated host, rather than an 
LG-like environment.

Here we introduce the first in a set of simulations that apply the FIRE  
physics to LG-like volumes at state-of-the-art resolution. We present two 
simulated LG-like pairs (containing 4 MW-mass analogues), along with six 
isolated MW-mass galaxies for comparison.  Our simulations generally 
reproduce the observed properties of dwarf galaxies in the LG:  they do not
suffer from either the missing satellites problem or TBTF when including baryonic physics.

This paper is organized as follows.  In \S~\ref{sec:sims}, we describe the
simulations and briefly review the star formation and feedback models.
\S~\ref{sec:catalogs} details our methods for compiling our observed and
simulated galaxy catalogs.  \S~\ref{ssec:massfuncs} presents the stellar mass
functions of our simulated hosts, counting both satellites and non-satellites.
\S~\ref{ssec:tbtf} then examines the internal structure of our simulated dwarfs
by comparing their central masses to  those implied by observations via circular
velocity curves. \S~\ref{ssec:scatters} presents the relationships between
stellar kinematics, stellar mass, and halo mass.  We summarize our results and
conclusions in \S~\ref{sec:conclusions}.

\section{Simulations}
\label{sec:sims}

\begin{figure*}
\centering
\includegraphics[width=0.9\textwidth]{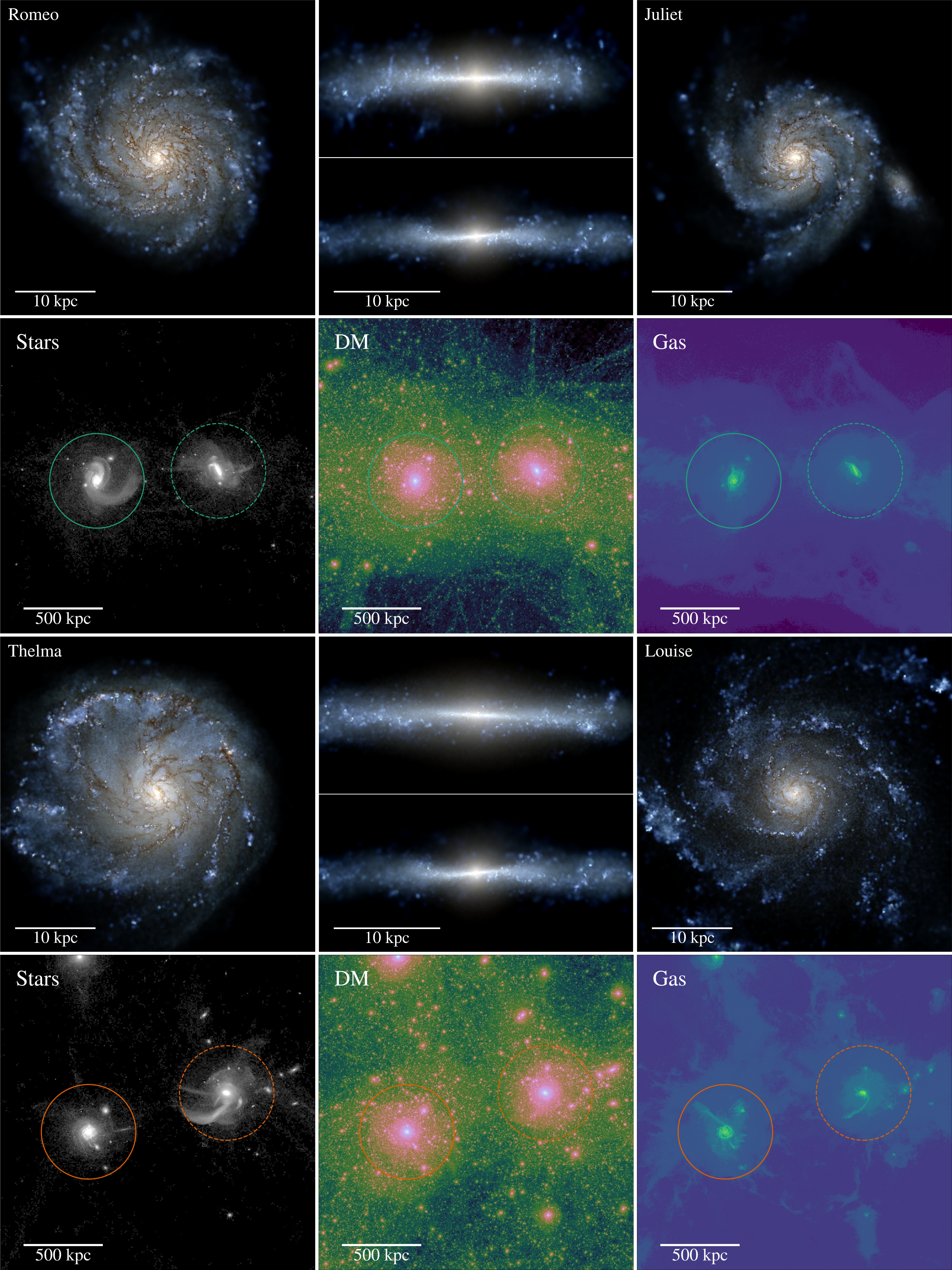}
\caption{Visualizations of our simulated hosts and their environments.
The face-on pseudo-color images are 40~kpc across; the edge-on images
span 30~kpc with a height of 15~kpc.  The density maps show the highest
3D density along a given line-of-sight through a cube 2~Mpc on a side,
centered on the mid-point of the pair.  All of the maps adopt logarithmic
color scales; the stellar maps range from
$10^{-9} \text{--} 3\times10^{-2}~M_\odot$~pc$^{-3}$, the dark matter from
$10^{-8} \text{--} 1~M_\odot$~pc$^{-3}$, and the gas from
$10^{-8} \text{--} 100~M_\odot$~pc$^{-3}$.  Circles around the hosts
indicate a radius of $300~\kpc$; the more massive host halo is on the
right and is indicated by a dashed circle.  The massive galaxy on the outskirts 
of \tal\ (with $\mvir=4.5\times10^{11}\msun$, $\mstar(<20~\kpc) = 1.58\times10^{10}\msun$) 
is $>1~\mpc$ from both hosts, excluding it from the analyses that follow.}
\label{fig:vizes}
\end{figure*}

\begin{table}
\begin{center}
\setlength\tabcolsep{4pt}%
\def\arraystretch{1.05}%
\begin{tabular}{lccccr}
Host         & $\mvir$          & $\mstar(<20~\kpc)$      & $\rcontam$ & $N_\mathrm{contam}$ & $N_\mathrm{MF}$, MW        \\
             & $[10^{12}\msun]$ & $[10^{10}\msun]$        & [kpc]      & ($<1~\mpc$)         & (DMO)                      \\
\hline\hline
\multicolumn{6}{c}{\textit{Paired hosts}}                                                              \\
M31          & $1.7\pm0.3^a$    & $10.3_{-1.7}^{+2.3\,b}$ & ---        & ---                 & ---    \\
Milky Way    & $1.3\pm0.3^c$    & $5\pm1^c$               & ---        & ---                 & ---    \\
\run{Romeo}  & 1.24             & 7.37                    & 514        & 4                   & 10 (7) \\
\run{Juliet} & 1.01             & 4.22                    & 1196       & 0                   & 15 (8) \\
\run{Thelma} & 1.32             & 7.92                    & 1215       & 0                   & 10 (6) \\
\run{Louise} & 1.03             & 2.86                    & 894        & 0                   & 8 (4)  \\
\multicolumn{6}{c}{\textit{Isolated hosts}} \\
\run{m12b}   & 1.31             & 9.42                    & 728        & 0                   & 9 (6)  \\
\run{m12c}   & 1.26             & 6.44                    & 1247       & 0                   & 14 (2) \\
\run{m12f}   & 1.54             & 8.79                    & 1110       & 0                   & 8 (5)  \\
\run{m12i}   & 1.07             & 7.00                    & 542        & 6                   & 13 (9) \\
\run{m12m}   & 1.45             & 12.62                   & 671        & 3                   & 15 (9) \\
\run{m12z}   & 0.80             & 2.24                    & 445        & 4                   & 7 (5)  \\
\end{tabular}
\end{center}
\begin{scriptsize}
$^a$ \citet{Diaz2014} \hfil
$^b$ \citet{Sick2015} \hfil
$^c$ \citet{BlandHawthorn2016}
\end{scriptsize}
\caption{Basic properties of our host halos: virial mass ($\mvir$, using the 
\citealp{Bryan1998} definition), stellar mass of the central galaxy ($\mstar(<20~\kpc)$), 
distance to the nearest low resolution particle ($\rcontam$), the number of halos 
within $1~\mpc$ with $\vmax\geq10~\kms$ that are excluded due to contamination 
from low resolution particles ($N_\mathrm{contam}$), and the number of massive 
failures identified when comparing subhalos in the corresponding DMO simulations 
with the MW dSphs (unaccounted for subhalos with $\vmax = 25-40~\kms$; see
\S~\ref{ssec:tbtf} for details); ``strong'' massive failures are given in
parentheses.  We caution that estimates for the virial masses of the MW and
M31 frequently vary at the factor of $\gtrsim2$ level \citep[e.g.][]{Kafle2018,
Patel2018}. Though we do not list it, the total fractional mass contamination
within $1~\mpc$ by low resolution particles is, at worst, $3.6\times10^{-4}$
around \run{m12i}.  Initial baryonic particle masses are $3,523~\msun$ in \raj,
$3,990~\msun$ in \tal, $4,174~\msun$ in \run{m12z}, and $7,067~\msun$ in the
remaining simulations.
}
\label{tab:info}
\end{table}

We analyze hydrodynamic, cosmological zoom-in \citep{Katz1993,Onorbe2014}
simulations, initialized with \texttt{MUSIC} \citep{MUSIC}, from the FIRE
project \citep{FIRE}, run using the improved ``FIRE-2'' version of the code from
\citet{FIRE2}.  All of the simulations were run using \texttt{GIZMO} 
\citep{GIZMO},\footnote{\url{http://www.tapir.caltech.edu/~phopkins/Site/GIZMO.html}} 
a multi-method gravity plus hydrodynamics code, in meshless finite-mass (``MFM'')
mode. This is a mesh-free Lagrangian finite-volume Godunov method which
automatically provides adaptive spatial resolution while maintaining
conservation of mass, energy, and momentum (for extensive tests, see
\citealt{GIZMO}). Gravity is solved with an improved version of the Tree-PM
solver from GADGET-3 \citep{Springel2005}, with fully-adaptive (and fully-conservative) 
gravitational force softenings for gas (so hydrodynamic and force softenings are 
always self-consistently matched), following \citet{Price2007}.

The FIRE physics and source code are nearly identical to those in previous
FIRE-2 simulations, with the lone exception that all of our simulations
additionally include subgrid turbulent metal diffusion, which produces more
realistic metallicity distributions in dwarf galaxies \citep{Escala2017} but
does not alter other galaxy-wide properties \citep{Hopkinsmetaldiff,Su2016}.
The FIRE physics modules are described in detail in the papers above, but in
brief, we treat radiative heating and cooling from $10-10^{10}\,$K, allow for
star formation only in gas that is dense ($n > 1000$ cm$^{-3}$), Jeans unstable,
molecular and self-shielding \citep{Krumholz2011}, and self-gravitating
\citep{Hopkins2013sf_criteria}.  We then include stellar feedback via radiation
pressure, Types Ia and II supernovae, metal mass loss, and photo-ionization and
photo-electric heating, assuming every star particle represents a single stellar
population with a \citet{Kroupa2001} IMF.

We focus on two pairs of LG-like hosts, \raj\ and \tal, which are visualized at
$z=0$ in Figure~\ref{fig:vizes}.  We refer to these simulations (and additional
ongoing work) as the ``ELVIS on FIRE'' set.  The \tal\ volume was first
presented as a DMO simulation as part of the original Exploring the Local Volume
In Simulations (ELVIS) suite \citep{ELVIS}.  Both \tal\ and \raj\ were also
presented at lower resolution and without subgrid metal diffusion in
\citet{m12morph}. We also include the results of six simulations targeting
isolated MW-mass halos; all of these galaxies were also analyzed in \citet{m12morph},
but here we present higher resolution resimulations of \run{m12b}, \run{m12c}, and
\run{m12z} that additionally include subgrid metal diffusion.
\run{m12b}--\run{m12m} are part of the ``Latte Suite,'' a set of hosts
homogeneously selected to be isolated and roughly the same mass as the
MW:  $M_{200_\mathrm{m}}=1-2\times10^{12}\msun$. \run{m12i}, in particular, 
uses the same initial conditions as the halo presented in \citet{Wetzel2016}, 
originally taken from the AGORA project \citep{Kim2014}.  The hosts in the Latte 
Suite were all simulated with identical resolutions:  initial baryonic particle 
masses $m_b = 7,067\msun$. Because the LG-like pairs were drawn from different 
box sizes and slightly different cosmologies,\footnote{%
All of our simulations assume flat $\Lambda$CDM cosmologies with $h = 0.68-0.71$, 
$\Omega_\mathrm{m} = 0.266-0.31$, $\Omega_\mathrm{b} = 0.0455-0.048$, and 
$\sigma_8 = 0.801-0.82$ \citep[e.g.][]{Larson2011,Planck15}. These slight differences 
in cosmology should have a negligible impact on the scale of the LG \citep[e.g.][]{GKBICEP}.%
}
they feature $\sim2\times$ better resolutions (\raj\ has $m_b = 3,523\msun$;
\tal\ has $m_b = 3,990\msun$). Finally, \run{m12z} was also chosen from a
separate parent box to be slightly lower mass, and is also at slightly higher
resolution than the remainder of the isolated sample with $m_b = 4,174\msun$.
All simulations were run with gas softening lengths that are fully adaptive 
down to $\egas \simeq 0.5-1~\pc$ and DM force softenings $\simeq50~\pc$.

The two central galaxies in \raj\ are separated by $839~\kpc$, are approaching
one another with $v_\mathrm{rad} = -93~\kms$, and have a tangential velocity of
$v_\mathrm{tan} = 23~\kms$.  \run{Thelma} and \run{Louise} are separated by
$920~\kpc$, have $v_\mathrm{rad} = -107~\kms$, and $v_\mathrm{tan} = 14~\kms$.
For comparison, the MW and M31 are separated by $787~\kpc$
\citep{McConnachie2005} and are approaching one another with $v_\mathrm{rad} =
-109~\kms$ and $v_\mathrm{tan} = 17\pm17~\kms$ \citep[][though see
\citealp{Salomon2016} and \citealp{Carlesi2016}]{vanderMarel2012}. Both pairs
were selected for these high resolution simulations on the basis of their low
tangential velocities and relative lack of (partial) overlap in their Lagrange
volumes with other massive halos outside the LG.  We do not constrain or 
restrict the larger-scale density fields around the LG hosts; i.e. we do
not necessarily expect to reproduce the $\sim5~\mpc$-scale ``Local Sheet''
\citep{McCall2014}.  Table~\ref{tab:info} presents additional information 
about the individual hosts, including the distance to the nearest low-resolution 
particle $\rcontam$ and the number of halos within $1~\mpc$ excluded from our 
analysis due to contamination from these particles.

\section{Galaxy catalogs}
\label{sec:catalogs}

In this section, we briefly discuss the observational sources we use for 
the properties of dwarf galaxies in the LG, along with our method for extracting
the equivalent properties for dwarf galaxies from the simulations.

\subsection{Observations}
\label{ssec:obscats}

We build our observational sample primarily off the data compiled in an updated
version of the \citet{McConnachie2012} catalog of local dwarf galaxies.  We
exclude all ``starred'' systems in the catalog, for which debate remains about
their true nature (i.e. galaxy \emph{vs.} globular cluster); the majority of
these are much less massive than our resolution.  We take stellar mass-to-light
ratios from \citet{Woo2008} where available, and otherwise assume
$\mstar/L_\mathrm{V} = 1.6$ (consistent with \citealp{Martin2008} and
extrapolations of \citealp{Bell2001}).  We calculate $\vhalf = \vcirc(\rhalf)$,
the implied circular velocity at the 3D (deprojected) half-light radius, for the
majority of our galaxies with the \citet{Wolf2010} formula, i.e. based on the
velocity dispersion of the stars.  For the MW dSphs, we use the velocity
dispersions presented in \citet{Wolf2010}.  For the satellites of M31, we take
$\rhalf$ and $\vhalf$ from \citet{Tollerud2014}. The majority of these are based
on stellar velocity dispersions, but there are a few exceptions.  Most notably,
the constraint on M33 only represents the mass of the dark matter halo, taken
from a fit to CO and HI observations (\citealp{Simon2006}, using data from
\citealp{Corbelli2000} and \citealp{Corbelli2003}); including the baryonic
component roughly doubles $\vhalf$.
We adopt the total mass estimates (i.e. including baryons) for the
remaining M31 satellites, including those that are baryon dominated within
$\rhalf$.  For NGC 185 and NGC 147, these are based upon the dynamical modeling
of \citet{Geha2010}, while the constraint on IC 10 is derived from HI
observations \citep{Wilcots1998}.  Finally, for the Local Field, we adopt
the values ($\rhalf$, $\vhalf$, and $\sigstar$) calculated or compiled in
\citet{Kirby2014} where possible, though we adopt the modified $\vhalf$ values
presented in \citet{ELVISTBTF} for the three galaxies that display evidence of
rotation:  for the dwarf galaxy WLM, we use the result calculated in detail by
\citet{Leaman2012}, while we use the method of \citet[][and also see
\citealp{Kirby2014}]{Weiner2006} to incorporate rotational support into our
estimates for Pegasus and Tucana.  For all other systems, we fall back on the
measurements in \citet{McConnachie2012}. We list the properties of the full
sample in Appendix~\ref{sec:obsgals}.

\subsection{Simulations}
\label{ssec:simcats}

Because publicly-available halo finders are typically tuned to capture DM
(sub)halos, we find unsatisfactory performance when attempting to capture the
much more compact stellar clumps (particularly when those clumps are embedded
within the stellar halo of a larger host; see Figure~\ref{fig:vizes}).  We
therefore compile our simulated galaxy catalogs via a multi-step process. We
first identify bound DM halos by running \texttt{AHF} \citep{AHF} only on the DM
particles.  We then assign star particles in a first pass to DM clumps via a
generous cut on stellar positions and velocities along the direction of motion
of the (sub)halo. In a second pass, stars are iteratively removed based on their
velocities relative to the velocity dispersion of the system until the latter
stabilizes. We then examine each galaxy by hand and repeat the final step with
a small maximum radius if necessary. Finally, we iteratively compute stellar
velocity dispersions independently along the $x$, $y$, and $z$ axes, eliminating
stars offset by more than $5\sigma$ from the mean until the dispersion along
each axis changes by less than $2.5\%$; this step typically alters particle
counts at the percent level. However, this step is important for velocity
dispersions because contamination by even a single background halo star, with
high relative velocity to the satellite, can significantly bias properties such
as the radius or velocity dispersion of the satellites.  We define $\mstar$ as
the sum of the masses of all the star particles that remain assigned to each
galaxy in this way and $\sigstar$ as the RMS average of the $x$, $y$, and $z$,
velocity dispersions of those particles (calculated via the interquartile
spacing).  Finally, we recompute $\vmax$ and $\rmax$, the radius at which
$\vmax$ occurs, using all particles around each host; this step is unimportant
for low mass galaxies, but matters in the higher stellar mass  dwarfs where the
star particles are a non-negligible fraction of the mass within
$\rmax$.\footnote{In cases where the circular velocity curve has no
peak/turnover, we instead adopt the inflection point of the curve, i.e. the
radius/circular velocity where the curve becomes convex due to the contribution
from a background host halo, as $\rmax$ and $\vmax$.}  We compute all properties
and profiles relative to a halo/galaxy center defined using a ``shrinking
spheres'' approach on the stars \citep{Power2003}.  Though there is no explicit
requirement at any step that star particles assigned to a given galaxy be bound
to the associated halo, our final velocity distributions suggest this is
typically the case.

Our approach is similar to \cite{Wetzel2016}, but we base our galaxy catalogs on
\texttt{AHF} halo catalogs (rather than \texttt{rockstar}; \citealp{rockstar}) and 
the cuts placed on stellar particles vary slightly; most notably, \citet{Wetzel2016} 
did not include either our initial cut based on the motion along the direction of the
subhalo or our final cut while computing velocity dispersions.  Moreover, we
quote total line-of-sight velocity dispersions, whereas \citet{Wetzel2016}
computed total velocity dispersions at the half-mass radius.  Our results are
similar: for example, we find an identical number of galaxies with
$\mstar\geq10^5\msun$ when applying our method to \run{m12i} as \citet{Wetzel2016}
identify in the same halo (simulated without metal diffusion).

In the figures that follow, we plot stellar mass functions down to 
$\mstar=7\times10^4\msun$, corresponding to approximately 10 star particles
in the lower resolution Latte simulations.  While the existence and stellar 
masses of galaxies above this cut is robust, the internal properties, such 
as density or velocity dispersion, are more sensitive to resolution and may 
change with higher resolution simulations \citep{FIRE2}.  We therefore adopt 
a slightly higher cut, $\mstar = 10^5\msun$, corresponding to $14-29$ star 
particles, when quoting galaxy counts or investigating internal structure.

\begin{figure*}
\centering
\includegraphics[width=0.33\textwidth]{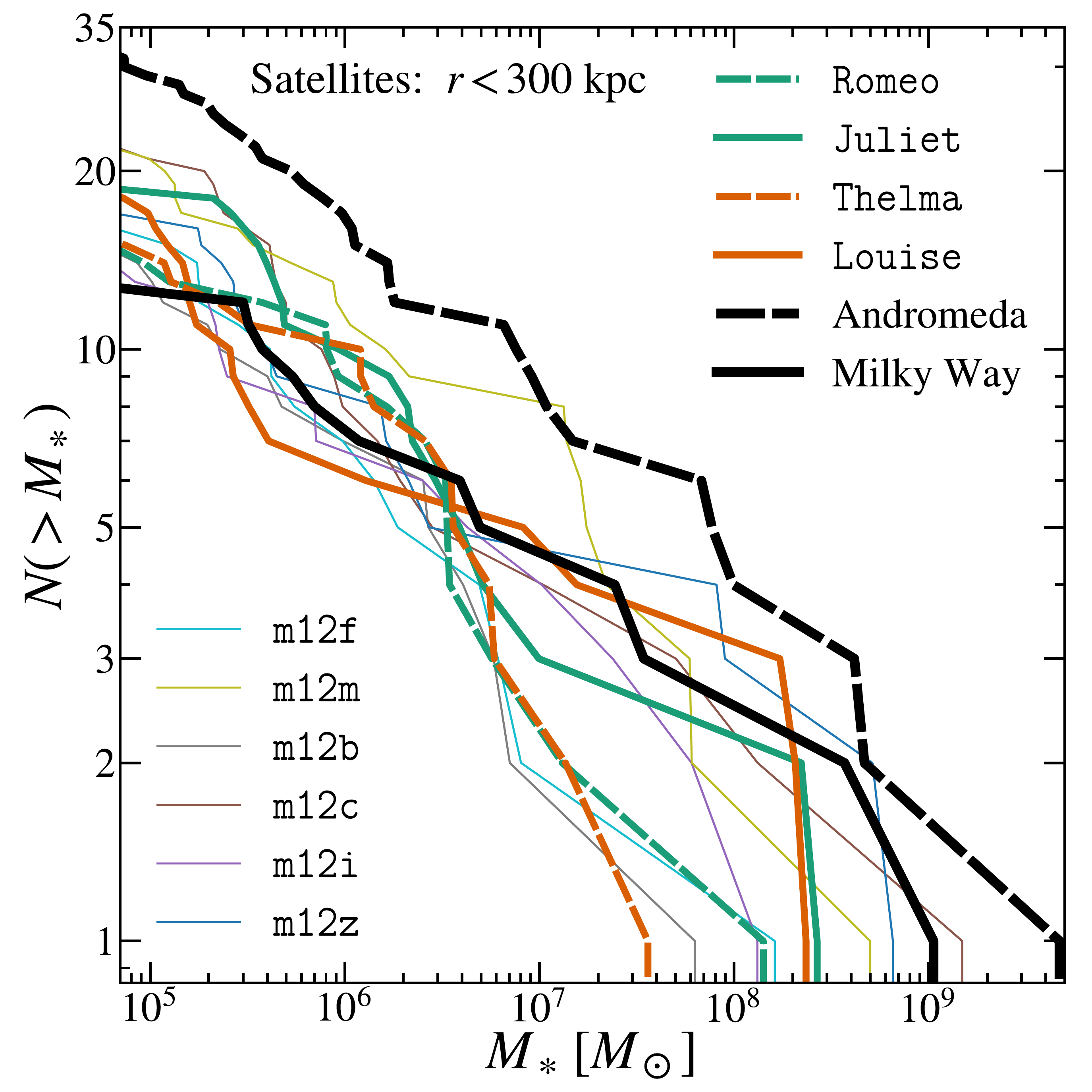}
\includegraphics[width=0.33\textwidth]{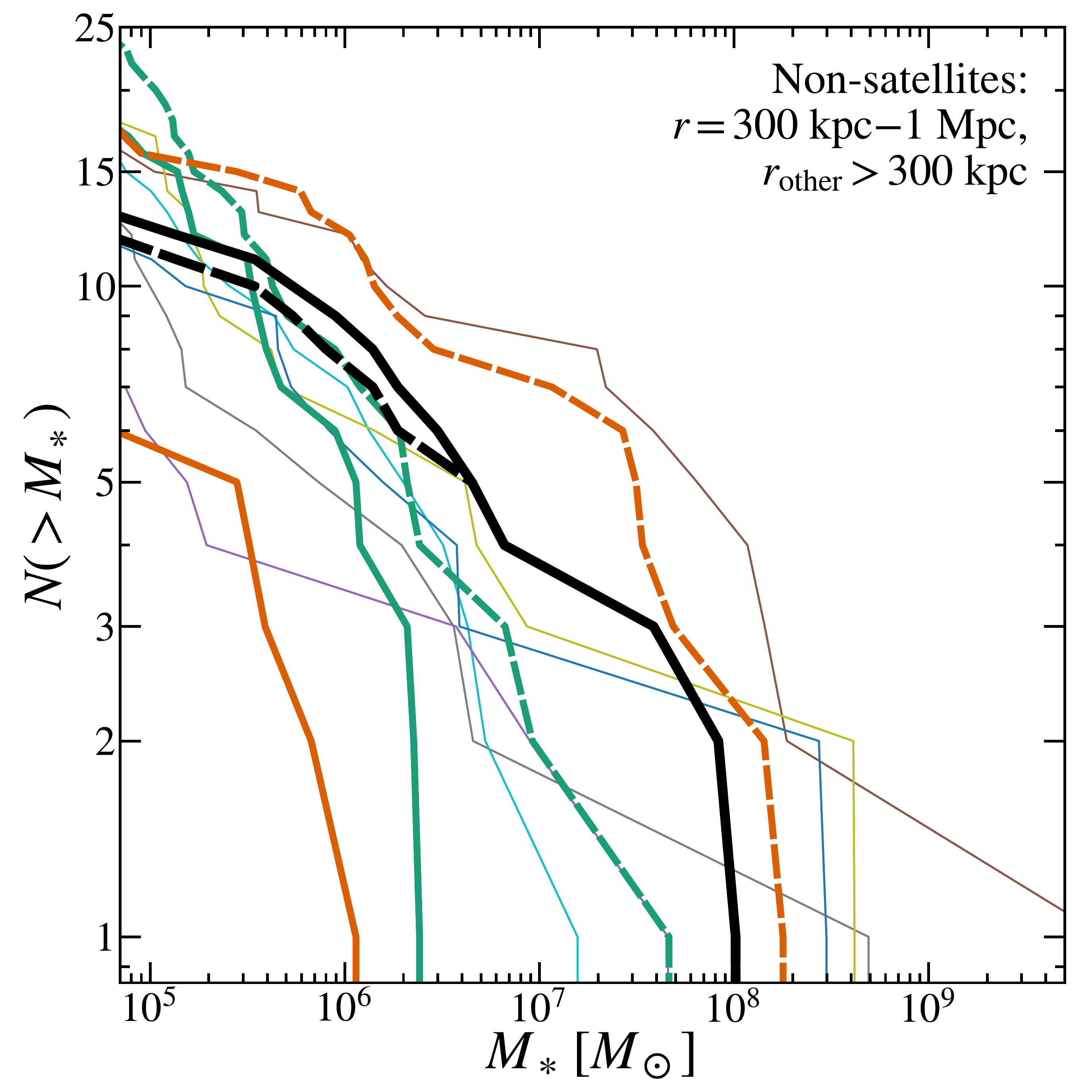}
\includegraphics[width=0.33\textwidth]{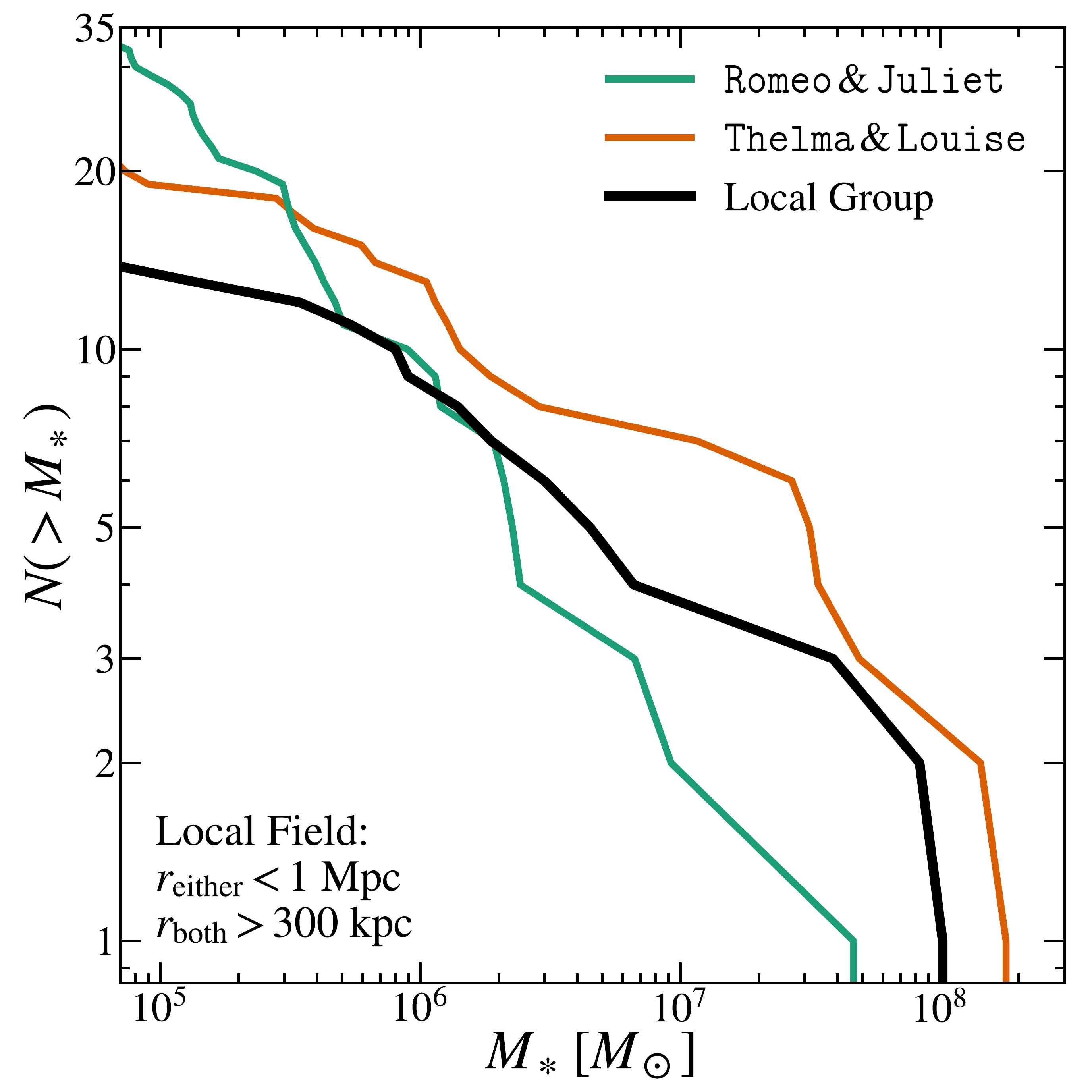}

\caption{Galaxy stellar mass functions. The panels indicate the satellite population 
(\textit{left}; host distance $r_\mathrm{host} < 300~\kpc$), the non-satellite population 
around each host (\textit{center}; $r_\mathrm{host} = 300-1000~\kpc$, and distance to 
the paired host $r_\mathrm{other} > 300~\kpc$ where applicable), and (\textit{right})
the Local Field (distance from either host $r_\mathrm{either} < 1~\mpc$ but distance from 
both hosts $r_\mathrm{both} > 300~\kpc$).  Thin lines indicate the isolated \run{m12}
sample, which are sorted in the legend by host virial mass.  The satellite stellar mass 
functions are broadly consistent with that of the MW and M31, though even our richest 
satellite populations slightly (by a factor of $\sim1.2$ at $10^5\msun$) under-produces 
that of M31, possibly because our highest mass host is only $1.45\times10^{12}\msun$.  
Similarly, the  non-satellite populations around each host are in reasonable agreement 
with that of the MW and M31, with considerable scatter.  
The simulated Local Field populations are also generally consistent with observations,
particularly for $\mstar \gtrsim 5\times10^5\msun$; below that, \raj\ displays a steep 
upturn relative the LG.  \tal, meanwhile, slightly overproduces the Local Field SMF at 
all masses.  We predict a median of 2.5 additional (i.e. undetected) non-satellite 
galaxies with $\mstar\geq10^5\msun$ and $r_\mathrm{MW} = 300-1000~\kpc$, along with 4 
additional MW satellites with $\mstar=10^5-3\times10^5\msun$.
}

\label{fig:mstar}
\end{figure*}

\section{Stellar mass functions}
\label{ssec:massfuncs}

Figure~\ref{fig:mstar} presents the stellar mass functions (SMFs) of  dwarf
galaxies throughout the Local Volume.  As expected from \citet{ELVIS},  the
satellite SMFs (host distance $r_\mathrm{host} < 300~\kpc$) of the isolated  and
paired halos overlap well.  Our ten hosts contain between $12$ and $20$ 
satellites with $\mstar\geq10^5\msun$, with a 66\% scatter of $6.1$ galaxies.  For 
comparison, the scatter in the number of subhalos around the DMO ELVIS hosts 
\citep{ELVIS} above an equivalent peak halo mass (using the zero-scatter 
stellar mass \emph{vs.} peak halo mass relationship from that work) is $20.5$.  
However, the host masses from ELVIS also vary more widely than the sample 
presented here:  the DMO ELVIS host masses have a 66\% scatter of $1.25\times10^{12}\msun$, while 
that of our sample is only $0.37\times10^{12}\msun$.  Naively scaling the
two values by one another (i.e. scatter in $N_\mathrm{sats}(\msun\geq10^5\msun)/$ 
scatter in host $\mvir$) yields nearly identical values, such that our results 
are consistent with the FIRE simulations predicting the same degree of scatter 
in the number of luminous satellites as DMO simulations.

The FIRE satellite populations also provide a good match to the MW satellite
SMF, particularly below the masses of the LMC and SMC,\footnote{The worse 
agreement at the high-mass end is not particularly unexpected:  none of our
hosts were selected to contain an LMC-mass satellite, and a randomly selected
MW/M31-mass halo is statistically unlikely to have LMC or M33-mass satellites
\citep{Busha2011,Tollerud2011}.} though the agreement is
not perfect: the simulated galaxies host a median of $15.5$ satellites with
$\mstar\geq10^5\msun$, compared with the $12$ such known MW satellites, and we
typically predict a SMF that continues to rise between the relatively bright
classical dSphs ($\mstar\gtrsim3\times10^5\msun$) and the ultra-faints dwarfs
($\mstar\lesssim3\times10^4\msun$) identified in deep surveys such as SEGUE
\citep{Belokurov2009} and DES \citep{DESDwarfs2015}.  The difference is small
relative to the order-of-magnitude difference referred to by the  missing
satellites problem -- we predict a median of 4 satellites with  $\mstar=10^5 -
3\times10^5\msun$ -- but it may suggest additional, relatively luminous,
undetected satellites \citep[also see][]{Tollerud2008}.  Rather than a sign of
observational incompleteness, the flattening of the MW SMF may instead reflect
a feature from reionization \citep[see][]{Bose2018}; if so, our simulations do
not capture such a feature overall.

In contrast to the relative agreement with the MW SMF, all of the simulated 
satellite SMFs lie slightly below that of M31.  Our hosts have, on average, $54\%$ as
many satellites with $\mstar\geq10^5\msun$ as are already known around M31.  The offset 
in the mean counts relative to M31 is roughly constant for $\mstar\lesssim10^7\msun$ (at 
which point the mean difference becomes even larger), indicating that M31 contains 
systematically more satellites at fixed stellar mass than our simulated hosts.  For 
comparison, the mean offset between the simulated satellite populations and that of the 
MW is $\sim2\%$ at the mass of CVnI ($3\times10^5\msun$) and remains under $20\%$ over 
two orders of magnitude (up to the mass of Fornax, $2.4\times10^7\msun$).  The difference 
in satellite counts is clear, but not extreme:  our host with the largest number of 
satellites (\run{m12m}, with $\mvir = 1.45\times10^{12}\msun$) contains $73\%$ as 
many galaxies above $10^5\msun$ with an average of $74\%$ from $10^5$ -- $3\times10^7$.  
As we show in Appendix~\ref{sec:mstar400}, this result is only marginally sensitive
to the radial cut used to separate satellites from non-satellites.  It is also
qualitatively independent of the assumed mass-to-light ratio for the observed
dwarf galaxies:  even adopting a stellar mass-to-light ratio of unity for the
galaxies not included in \citet{Woo2008} yields a mean of $61\%$ as many 
satellites as M31 with $\mstar=10^5\msun$.

\begin{figure} 
\includegraphics[width=\columnwidth]{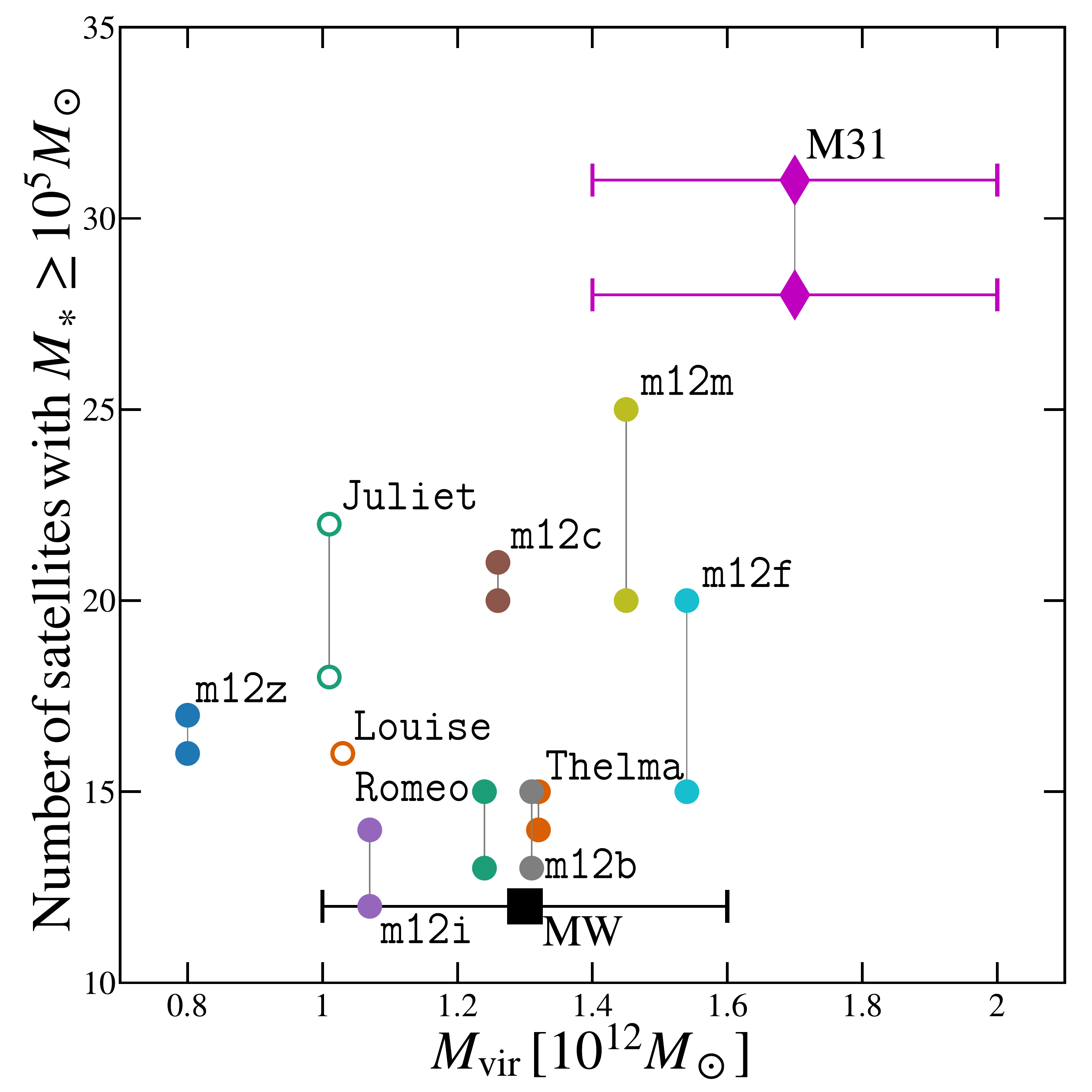}
\caption{The number of dwarf galaxies with $\mstar\geq10^5\msun$ within
$300~\kpc$ (lower points) and $400~\kpc$ (upper points) of each host, as a
function of host virial mass.  Colors are  identical to Figure~\ref{fig:mstar},
with the lower mass host in the LG-like pairs plotted as open points.  Counts
around M31  and the MW are also plotted, with mass estimates taken from
\citet{Diaz2014} and \citet{BlandHawthorn2016}, respectively. Both the MW and
\run{Louise} have zero satellites with $\mstar\geq10^5\msun$ between
$300-400~\kpc$ (Samuel et al., in preparation), and therefore have only 
a single value plotted.}
\label{fig:nsats} 
\end{figure}

The abundance of dwarf galaxies around M31 (relative both to the MW and to our 
simulated hosts) may point towards a higher M31 halo mass.  Large-scale estimates 
for the mass of M31 typically suggest $M_\mathrm{vir,\,M31}\gtrsim1.5\times10^{12}\msun$; 
for example, \citealt{Diaz2014} used the net momentum of the LG to estimate 
$M_\mathrm{vir,\,M31} = 1.7\pm0.3\times10^{12}\msun$.  However, \citet{Kafle2018} 
recently argued for $M_\mathrm{vir,\,M31}=0.8\pm0.1\times10^{12}\msun$ by applying 
a Bayesian framework to high-velocity planetary nebulae. Figure~\ref{fig:nsats}
shows the number of dwarf galaxies near each host, as a function of host virial 
mass. Though the trends with mass are weak (e.g. our lowest mass host 
contains the fifth most satellites), our results suggest that it is difficult to 
match both the SMF of the MW and of M31 without a higher virial mass for M31.  

Broadly speaking, the non-satellite SMFs in Figure~\ref{fig:mstar} ($r_\mathrm{host} = 300-1000~\kpc$, and
excluding satellites of the paired host if applicable) generally agree with
counts in the fields around the MW/M31.  However, there are again hints of
undetected galaxies with $\mstar\gtrsim10^5\msun$: we predict a median of $14.5$
galaxies with $\mstar\geq10^5\msun$, compared to the $12$ known around the MW.
Furthermore, increasing the mass of our M31 analogue may result in even more
predicted dwarfs; our predictions in the Local Field may be a lower limit. If
ultra-diffuse galaxies (UDGs) are prevalent in the field (as predicted by
\citealp{DiCintio2017} and \citealp{Chan2017}), with central surface
brightnesses $24-26~\mathrm{mag\,arcsec^{-2}}$ \citep{vanDokkum2015}, then some
of this incompleteness may even arise at $\mstar\sim10^7\msun$.  Surprisingly
(as \citealp{ELVIS} predict $75\%$ more halos above fixed $\vmax$ in DMO halo
counts), there is no clear offset in the Local Field SMFs between the isolated
and paired hosts, though all of the latter except \run{Louise} are on the upper
edge of the distribution.  However, our statistics remain relatively small, and
we require a larger, mass-selected sample to make strong statements regarding
the efficiency of galaxy formation in dwarfs within $\sim1~\mpc$ of an LG-like
pair \emph{vs.} an isolated MW-mass galaxy.  We caution that the lines
representing \run{Romeo} and \run{Juliet} (\run{Thelma} and \run{Louise}) are
not completely independent, with the volumes probed overlapping by 42\% (37\%).

Finally, the right panel of Figure~\ref{fig:mstar} plots the SMF of the ``Local
Field'' (all non-satellite galaxies within $1~\mpc$ of either of the hosts). The
observed Local Field SMF lies roughly in between our two simulated LGs for
$\mstar\gtrsim5\times10^5\msun$.  Consistent with the center panel, some amount
of  observational incompleteness is possible, and perhaps even likely, but more
simulations  are required: both the real Local Field and \tal\ contain 5
non-satellite galaxies  with $\mstar=10^5-10^6\msun$, while \raj\ contains 19.
\tal, however, does  overproduce the observed SMF at all masses, predicting a
total of 18  galaxies with $\mstar>10^5\msun$ compared to the only 13 known in
the LG. However the comparison with the field around our larger sample of isolated hosts
clearly demonstrates very large systematic halo-to-halo variations in this prediction.

In Appendix~\ref{sec:mstar400} we consider the effects of a slightly larger 
($\sim 400\,$kpc) radial cut used to assign satellites their hosts, and show 
this does not qualitatively alter our conclusions above. However, it somewhat 
decreases the tension with both M31 and the Local Field by re-assigning 
a few galaxies from the field to the M31-analogue.


\section{Too-big-to-fail (TBTF)}
\label{ssec:tbtf}

Due to the resolution required to study the inner $\sim500~\pc$ of simulated
dwarf subhalos, TBTF was originally defined using DMO simulations. \cite{BK2011}
therefore focused on the dSph satellites of the MW.  Because dSphs are
dispersion supported, a measurement of $\sigstar$ provides a robust estimate of
$\vhalf$.  Moreover, the high dynamical mass-to-light ratios implied by
$\sigstar$ suggest that dSphs are strongly DM-dominated, indicating that the
estimates on $\vhalf$ may be fairly compared to the subhalo masses provided by
DMO simulations.  Later work on TBTF that expanded beyond the MW satellites
\citep[e.g.][]{ELVISTBTF,Tollerud2014} typically sought to recast observational
measurements for non-dispersion supported systems into similar constraints on
$\vhalf$, and either excluded or treated separately galaxies with significant
baryonic mass within $\rhalf$ (for which $\vhalf$ is not fairly comparable to the
results of DMO simulations).

Approaches to TBTF using baryonic simulations have varied.  For example,
\citet{Sawala2016} showed that the number of luminous subhalos in the APOSTLE
simulations above a given $\vmax$ agree with estimates for the MW satellite
population from \citet{Penarrubia2008}. They then obtain separate $\vmax$
estimates for the MW satellites by matching them with dwarf galaxies in their
simulations based on $\mstar$, $\vhalf$, and $\rhalf$; the $\vmax-\mstar$
relationship implied by these estimates is in good agreement with the simulated
relationship.  \citet{Wetzel2016}, conversely, sought to compare directly with
the data:  they showed good agreement between the dwarf satellites of \run{m12i}
and those of the MW/M31 when counting galaxies by stellar velocity dispersion
and when viewed in velocity dispersion -- stellar mass space.

Here, we adopt a hybrid approach.  We first demonstrate that the DMO simulations
of our host halos suffer from TBTF by reproducing the \citet{ELVISTBTF} analysis
on the DMO simulations, then show that the same analysis applied to the luminous
dwarf galaxies in the FIRE simulations yields no such discrepancy.  Because
direct comparisons with data are ideal, we will demonstrate in
\S~\ref{ssec:scatters} that the simulated dwarfs also broadly reproduce the
observed relationship between stellar mass and stellar velocity dispersion.
However, because we will compare our simulated dwarfs to non-satellite galaxies 
and to more massive systems, for which the assumption of dispersion-dominated
kinematics is not well-motivated, we begin by inspecting the central masses
of our simulated systems and their observational counterparts.

We therefore begin by generally replicating the analyses of \citet{BK2012,
Tollerud2012}, and \citet{ELVISTBTF}, who identified problematic (sub)halos by
comparing the circular velocity curves of simulated systems with constraints on
observed dwarf galaxies.  Before presenting the results of this analysis, we
first describe our methods for calculating the rotation curves in the DMO and
hydrodynamic simulations, then briefly review the galaxies included on each
plot, and finally summarize our nomenclature and methods for identifying and
counting the problematic (sub)halos.

\subsection{Methods}
\subsubsection{Calculating circular velocity curves}
For the DMO simulations, we follow previous TBTF analyses in computing circular
velocity curves for the (sub)halos by normalizing a fixed density profile to the
large-scale properties of each system.  We assume NFW \citep{Navarro1996}
profiles for the DMO systems, scaled to $\rmax$ and $\vmax$ of each halo, but,
as we discuss in \S~\ref{sssec:caveats}, this has a second-order effect on our
conclusions --  adopting the raw particle data from the DMO simulations and
ignoring the impact of  gravitational softening does not alter our conclusions.

Meanwhile, for the hydrodynamic simulations, we follow \citet{BK2012} in fitting
density profiles (here taken to be the $\alpha, \beta, \gamma$ model; e.g.
\citealp{Jaffe1983,Hernquist1990,Merritt2006} or \citealp{DiCintio2014}) to the
resolved portion of each halo, then extrapolating the fits inward to compute
$\vcirc(r)$.  Based on \S~4.1.4 of \citet{FIRE2}, who argued that the usual
\citet{Power2003} relaxation time criterion is equivalent to a limit on the
number of enclosed particles, we take $r_\mathrm{min}$ (the minimum radius used
in fitting the density and the radius within which we adopt the extrapolated
$M(r)$) as the radius containing 300 DM particles; we adopt an outer radius for
the fit of $15~\kpc$. Appendix~\ref{sec:fitting} directly examines the (minimal)
impact of varying $r_\mathrm{min}$, and compares $\vcirc$ from the extrapolated
fits to the raw data and to NFW profiles.  Importantly, as with the DMO
simulations, we show in \S~\ref{sssec:caveats} that the shape of the central
profile has only a marginal impact on the number of massive failures that we
identify in the hydrodynamic simulations, even among non-satellite galaxies.

\subsubsection{Selecting galaxies and halos} We separately analyze satellites of
the MW, satellites of M31, and galaxies in the Local Field, where satellites are
again defined as galaxies within $300~\kpc$ of each host.  We include every
galaxy that meets each distance cut and has velocity information that is
representative of the mass of the galaxy. This breaks slightly from the analyses
of \citet{BK2011,BK2012} and \citet{ELVISTBTF}, who eliminated the LMC, the SMC,
and NGC~6822 for various reasons.  In contrast, we eliminate only the
Sagittarius dSph.  Because Sagittarius is in the process of tidally disrupting,
stellar kinematics do not necessarily probe the underlying dynamical mass.
Consequently, we may identify a single subhalo as a ``massive failure'' (defined
in detail in \S\ref{sssec:massivefails}) that could be associated with
Sagittarius, increasing our counts below by one.  We generally adopt the 
constraints at ($\rhalf$, $\vhalf$) detailed in \S~\ref{ssec:obscats}, but the 
wealth of data on the Magellanic Clouds allows us to plot rotation curves for those
systems. Specifically, we adopt the HI-based rotation curve for the SMC from
\citet{Stanimirovic2004} and the proper motion-based rotation curve for the LMC
from \citet{vanderMarel2014}.  Finally, we note that M32 lies outside the limits
of the central panel (in the upper left, at $\rhalf = 110~\pc$, $\vhalf =
79~\kms$), and Leo T lies outside the limits of the right panel (at $\rhalf =
152~\pc$, $\vhalf = 13~\kms$).  Though these points are not shown on the axes,
they are included when identifying massive failures.

For the DMO simulations, we seek to reproduce the cuts adopted by previous TBTF
analyses.  However, because we lack evolutionary histories for our (sub)halos,
we select on present day $\vmax$ instead of adopting the
$\mathrm{max}\left[\vmax(t)\right] > 30~\kms$ cut used in, e.g.
\citet{ELVISTBTF}.  Based on Figure~1 of \citet{BK2012} and the results of
\citet{ELVISTBTF}, we consider (sub)halos with $\vmax\geq25~\kms$.  For
satellites, this cut is typically more conservative than the criteria of
\citet{ELVISTBTF} as many subhalos that reached $\vmax \gg 30~\kms$ can be
stripped to $\vmax\leq20~\kms$ today \citep[e.g.][]{Sawala2014}.  In principle,
however, we may include some systems (particularly in the Local Field) that only
recently reached their present day mass, and which may therefore be expected to
remain ``dark'' \citep[e.g.][]{Fitts2017}. However, as we will show below, there
are enough systems with $\vmax\geq25~\kms$ in the field that this is unlikely to
change our conclusions.\footnote{Our results with respect to the DMO simulations 
are insensitive to these cuts.  For example, we find qualitatively identical 
results if we select potential massive failures by their circular velocity at fixed
radius, rather than by $\vmax$.  Specifically, selecting the twelve subhalos with 
the largest circular velocities at $r=1~\kpc$, rather than all subhalos with 
$\vmax\geq25~\kms$, still yields at least one, and typically $\gtrsim3$, satellites 
with $\vcirc$ profiles that are incompatible with all of the MW dSphs (i.e., 
massive failures).}

For the hydrodynamic simulations, we opt to reproduce the cuts placed on the
observed galaxies.  That is, we select galaxies based on $\mstar$, rather than
$\vmax$.\footnote{Note, however, that we do assign galaxies to host the LMC and
SMC based on their $\vmax$, rather than $\mstar$, which is a more stringent cut
(see Figure~\ref{fig:scatters}).}  We select all luminous galaxies with
$\mstar\geq10^5\msun$.  As we show explicitly in \S~\ref{ssec:scatters}, this
cut is less restrictive than a $\vmax$-based cut:  it includes many halos with
$\vmax\ll25~\kms$, and only excludes three with $\vmax\geq25~\kms$.  Based on
Figure~\ref{fig:mstar}, this is a conservative estimate for a stellar mass-based
cut: the simulations all match or slightly exceed the MW SMF at $10^5\msun$.
The same is true in the Local Field: while observational completeness in the
Local Field is poorly defined, Figure~\ref{fig:mstar} shows that there are
likely undetected galaxies at $\mstar\lesssim5\times10^5\msun$.

\begin{figure*}
\centering
\includegraphics[width=\textwidth]{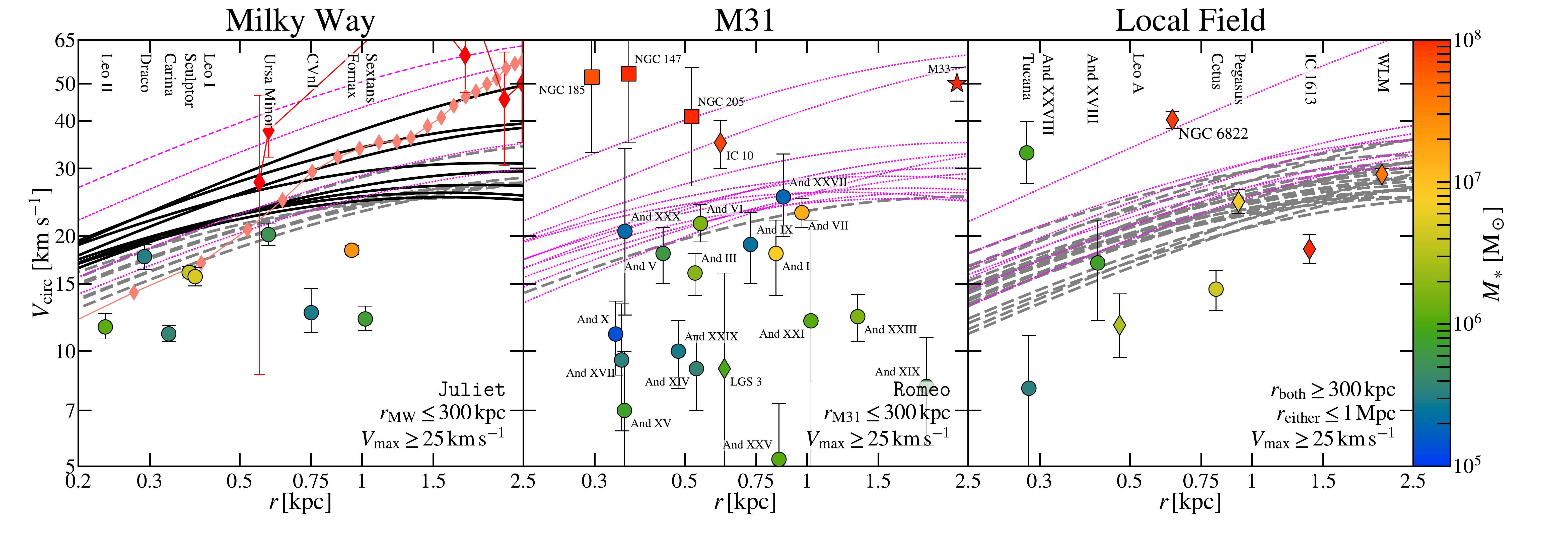} \\
\includegraphics[width=\textwidth,trim=0 0 0 3em]{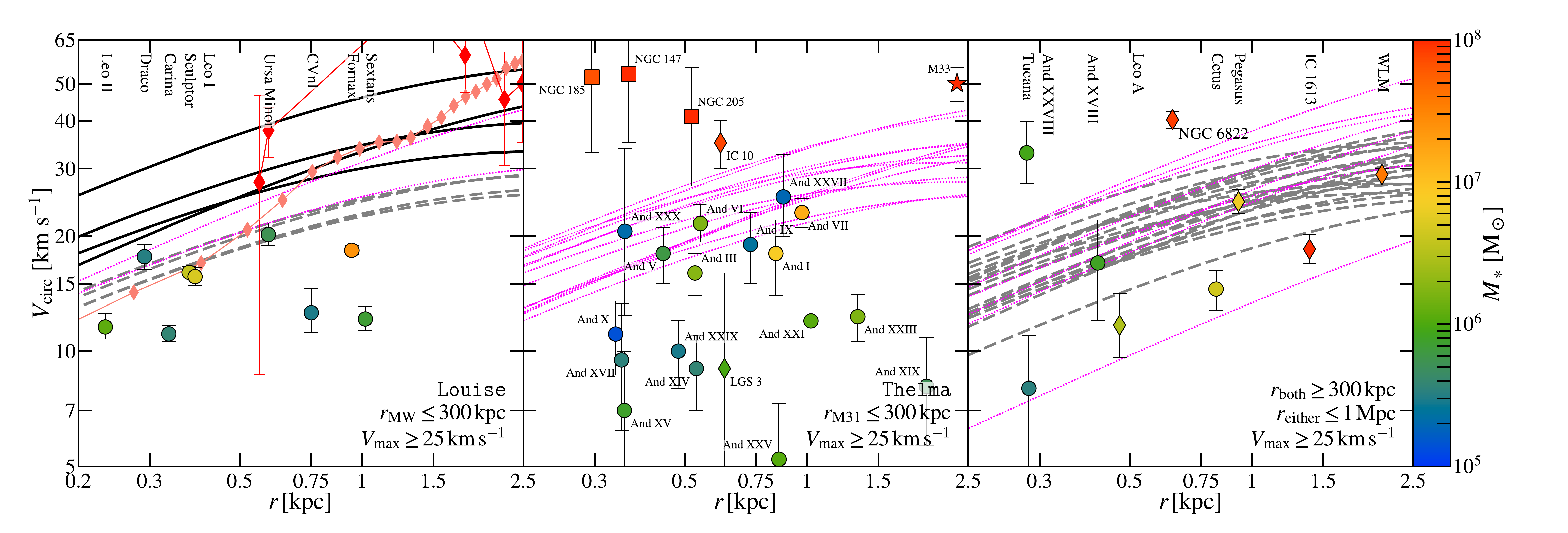}
\caption{
Circular velocity curves of dwarf (sub)halos in the DMO simulations, selected
according to $\vmax$, throughout \raj\ (top) and \tal\ (bottom). From left to
right, the panels plot MW satellites, M31 satellites, and galaxies in the Local
Field.  Circles, squares, and diamonds represent dSphs, dEs, and dIrrs,
respectively, with galaxy classifications taken from the literature; the star
indicates M33, and the lines marked with diamonds indicate rotation curves for
the SMC (small diamonds) and the LMC (large diamonds).  ``Strong'' massive
failures, which are halos too dense to host any of the galaxies in the
comparison sample other than the LMC and SMC, are plotted as solid black lines.
The less stringently defined massive failures, which are halos expected to host
relatively bright galaxies but that lack an observational counterpart, are
plotted as dashed grey lines. Halos assigned to host a galaxy are plotted in
magenta. The subhalos assigned to host the LMC and SMC (defined to be the two
most massive, if they have $\vmax\geq65$~and~$60~\kms$ respectively) are plotted
as short and long dashed magenta lines around \run{Juliet}. Both the M31 and the
Local Field contain dwarfs that are dense enough to eliminate all strong massive
failures and, when the dEs and M32 (outside the plot axes) are accounted for,
typically only a few subhalos with $\vmax\geq25~\kms$ remain unaccounted for
around M31. However, the TBTF problem, as identified by \citet{BK2011,BK2012}
around the MW and by \citet{ELVISTBTF} in the Local Field, exists in the DMO
simulations of all of our systems.  Every host has several subhalos that are too
dense to host any of the MW dSphs, along with many more that are only consistent
with Draco and Ursa Minor, and every Local Field analogue contains a plethora of
massive subhalos, many of which can only be associated with either Tucana or the
baryon dominated NGC 6822.}
\label{fig:vcirc_dmo}
\end{figure*}

\subsubsection{Identifying massive failures}
\label{sssec:massivefails}
We adopt the nomenclature of \citet{ELVISTBTF} in defining ``strong massive
failures'' and ``massive failures'' separately. Around the MW, the former are
subhalos that are too dense to host \emph{any} of the MW dSphs, while the latter
have rotation curves consistent with either Draco or Ursa Minor (or both), but
cannot be associated with those galaxies because they have already been assigned
to other subhalos.  In other words, strong massive failures have circular
velocity curves that lie above \emph{all} of the MW dSphs, while massive
failures are ``leftover'' systems that are otherwise consistent with either
Draco or Ursa Minor, but that are kinematically incompatible with the remainder
of the MW dSphs.

Due to the wide variability in the internal structures of dwarfs around M31 and
in the Local Field, we opt to apply the same nomenclature to those volumes but
insist that every galaxy be associated with a single halo (rather than just
Draco and Ursa Minor).  In practice, we therefore identify massive,
unaccounted-for halos.  As demonstrated by \citet{ELVISTBTF}, applying a stellar
mass \emph{vs.} halo mass relationship that reproduces counts in the Local Group
(when applied to DMO simulations) to these unaccounted-for halos assigns them
$\mstar\geq5\times10^5\msun$. Therefore, the massive failures we identify around
M31 and in the Local Field would be nominally expected to host bright galaxies.

\subsection{Results:  dark matter-only simulations}
Figure~\ref{fig:vcirc_dmo} presents the results of performing these analyses on
the DMO simulations. We compare the satellites of the lower (higher) mass host
in each pair to those of the MW (M31) in the left (central) panel, and show the
Local Field population in the right panel.  Strong massive failures (which
only exist in comparison with the MW satellites) are plotted as black lines,
while massive failures are indicated by the dashed grey lines. These latter set
are massive, dense (sub)halos that we nominally expect to form stars, yet which
lack an observational counterpart. Halos assigned to host a galaxy (which are
not counted as massive failures) are indicated by magenta lines. \run{Juliet}
contains analogues for both the LMC and SMC; these subhalos are indicated in the
long and short dashed magenta lines, respectively.

\begin{figure*}
\includegraphics[width=\textwidth]{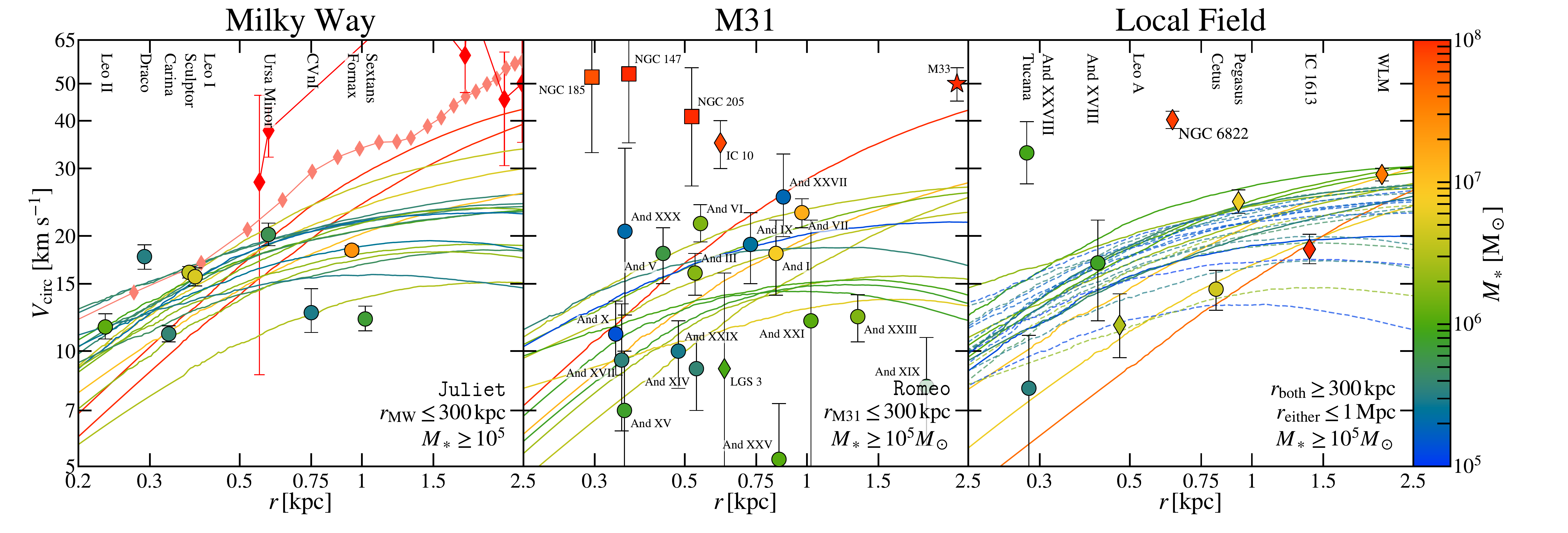} \\
\includegraphics[width=\textwidth,trim=0 0 0 3em]{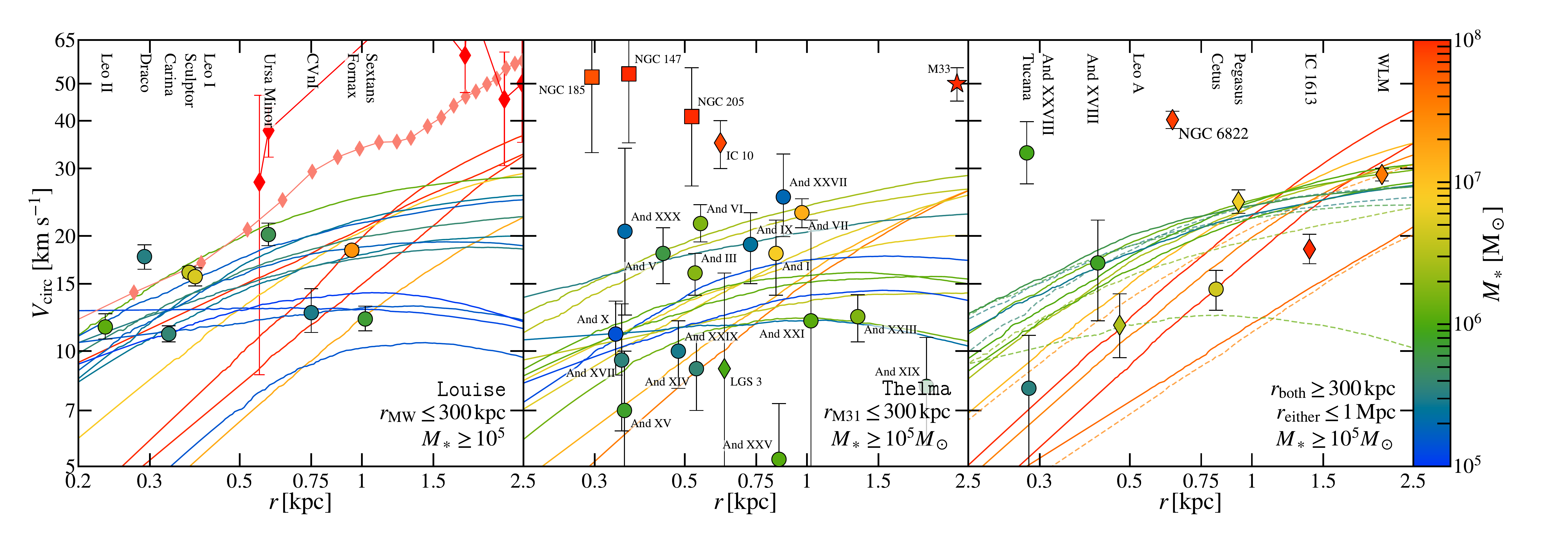}
\caption{Identical to Figure~\ref{fig:vcirc_dmo}, but here plotting $\vcirc$
curves from the hydrodynamic simulations.  Including baryonic physics using the
FIRE models eliminates TBTF around the MW and M31.  The dotted lines in the
Local Field panel show the persistence of several ``failures''
unaccounted for by current data, but these are quite different from the massive failures in the DMO runs: they have 
rotation curves similar to the typical observed LG and Local Field systems (there 
are simply $\sim10$ more of them). The mis-match may therefore be a result of observational incompleteness at $\mstar\lesssim10^6\msun$. The simulations here do not produce any galaxies with densities as \emph{high} as those of the baryon-dominated compact dEs around M31 (or Tucana/NGC 6822), with $V_{\rm circ} \gtrsim 35\,\kms$ at $r<1\,$kpc.}
\label{fig:vcirc_hydro}
\end{figure*}

As expected, we identify several (strong) massive failures in the left panel.  
However, our analysis identifies only one massive failure when comparing
\run{Romeo} to the M31 satellite population, and none among the satellites of
\run{Thelma}, though our analysis places several galaxies in subhalos that are
likely \emph{not massive enough} to host them.  As a glaring example, none of
the satellites of \run{Thelma} have $\vmax\geq50~\kms$, but four are assigned to
host M33, M32, NGC 205, and NGC 147, all of which have $\mstar\gtrsim10^8\msun$.
Moreover, our criteria identifies massive failures (relative to the M31
satellites) in \run{Juliet} (7) and in several of the isolated hosts: \run{m12c}
contains 6, \run{m12i} contains 3, and \run{m12m} contains 7.  We also remind
the reader that the hydrodynamic versions of these halos underproduce the SMFs;
if this is due to the masses of our hosts, then we would expect to also
underproduce the halo mass function, which scales closely with host mass
\citep[e.g.][]{Boylan-Kolchin2010}. Finally, both pairs contain a glut of
unaccounted for, massive halos in their Local Field populations. Moreover, in
both pairs, at least two of those leftover halos are too dense to be associated
with any of the known galaxies other than Tucana or NGC 6822.

We emphasize that all of our DMO hosts suffer from TBTF (as formulated by
\citealp{ELVISTBTF}) when comparing their satellite populations with the
satellites of the MW.  Though we only directly plot \run{Juliet} and
\run{Louise} against the MW satellites, we list the number of massive failures
(and, in parentheses, strong massive failures) in the final column of
Table~\ref{tab:info}:  in the DMO simulations, all of our hosts contain at least
two strong massive failures.

\subsection{Results:  FIRE simulations}
Figure~\ref{fig:vcirc_hydro} is analogous to Figure~\ref{fig:vcirc_dmo}, but it
plots $\vcirc$ curves of the luminous galaxies in the FIRE simulations (i.e.
including baryons).  Because we color the lines by stellar mass, we separate
massive failures and halos that are matched with observed dwarfs via line-style:
massive failures are plotted with dashed lines and the halos assigned to host
galaxies with solid lines. The addition of baryonic physics to the simulations
eliminates the TBTF problem around the MW and M31.  In particular there are
neither `strong massive failures' nor `massive failures' within the virial
radius of either host according to the definitions applied to the DMO simulations 
above. While the M31 population looks good in comparison to the TBTF
problem,  our hosts do not contain quite as many satellites as M31 overall: matching
the stellar mass function may result in additional galaxies that cannot be
matched one-to-one with observed systems.

There do remain a number of ``failures,'' according to our formal definition in 
the Local Field population (dotted lines), all with stellar masses $<10^{6}\msun$.  
However, we emphasize their circular velocities are still much lower than in the 
DMO simulations; in fact, they have profiles quite similar to the typical observed 
systems in both the MW, M31, and Local Field. Given that the completeness of the 
Local Field out to $\sim$\,Mpc at these masses is rather uncertain, one possibility 
is that there is a population of $\sim 10$ undetected dwarf galaxies in this region, 
with stellar masses $\mstar = 10^{5-6}\msun$ and dark matter densities similar to 
those of known dwarf galaxies (e.g.\ And XVIII).\footnote{Specifically, there are 17 (7)
of these missing systems in the Local Field of \raj\ (\tal) with $\mstar>10^5\msun$ 
and 7 (6) with $\mstar> 3\times10^5\msun$.}  However, we also note that this tension, 
like that in the Local Field stellar mass function, can be reduced (decreasing 
the number of discrepant halos by a few), without introducing significant tension 
in the comparison with TBTF around M31, if we use a larger radial cut as in 
Appendix~\ref{sec:mstar400} to associate galaxies with M31 and the MW.

Note that the relative impact of supernovae feedback is such that more massive 
dwarfs ($\mstar\sim10^8\msun$) almost universally have lower \emph{central} masses 
than their less luminous counterparts ($\mstar\lesssim10^6\msun$), particularly 
in the Local Field. Measuring dynamical masses within $\sim500~\pc$ across a 
range of stellar masses (e.g. with thirty meter-class telescopes) will test this 
prediction.

Aside from the the Local Field, our hydrodynamic simulations are free of TBTF:
all of the simulated dwarf satellites are consistent with even the lower density
MW dSphs and the satellites of M31. As we will show more quantitatively in
\S~\ref{ssec:scatters}, the stellar kinematics of the simulated galaxies are
also in line with those of dwarfs throughout the LG.

The agreement between the central masses of the simulated and observed galaxies
is not perfect, however:  the satellite populations do not contain any systems
quite as dense as NGC 205, NGC 147, NGC 185, or IC 10.\footnote{They also do not
contain any as dense as M32, but the high density of M32 may be at least
partially explained by a nuclear supermassive black hole
\citep{vanderMarel1998}, which we do not model in these simulations.}  This
result holds across our entire sample:  none of our hosts have satellites (or
field galaxies) that reach even the lower $1\sigma$ error on NGC 205, the least
dense of the dEs. Though this may be due to a lack of high mass dwarf galaxies,
the trend is typically in the opposite direction, such that our high mass dwarf
galaxies have relatively low $\vcirc$ at $\sim300~\pc$.  An examination of
Figure~6 of \citet{Sawala2016} and Figure~3 of \citet{Dutton2016} suggests that
the APOSTLE and NIHAO simulations, respectively, may also lack analogues of the
high density M31 satellites (halos with $\vcirc\sim50~\kms$ at $\sim500~\pc$).
These high density galaxies may represent a manifestation of the ``diversity
problem'' \citep{Oman2015,Creasey2017} in the LG.

Producing such high density galaxies, with $\mstar\sim10^8\msun$, may prove to
be an important test of galaxy formation physics.  In particular, while abundance
matching arguments suggest that these galaxies are at the centers of halos that
reached $\sim10^{10.5}\msun$ \citep{ELVIS}, previous work has shown that mass
scale to be the \emph{most} susceptible to core formation and stellar migration
due to supernovae feedback \citep{DiCintio2014,Chan2015,ElBadry2016}.  Some of
these could be the stripped cores of previously more massive galaxies: for
example, \citet{McConnachie2004} identified a stream that is likely originating
from  NGC~205.  However, they estimate the total mass in that stream to be only
$\sim2.5\%$ of the mass of NGC~205.  Moreover, this option is unlikely for at
least IC 10, which is gas rich and star forming today.  Furthermore, the
galaxies in the LG that are more massive than this sample, the LMC and M33, lack
these high density central clumps.  An additional, constant source of feedback
\citep[e.g. cosmic rays;][]{Jubelgas2008} that acts to smooth out the burstiness
in the star formation, leading to less-violent feedback episodes, may be required to
explain these objects.  For a more detailed discussion of the structure of
\emph{isolated} galaxies at this mass scale in the FIRE-2 simulations, we refer
the reader to \citet{Chan2017}, who studied the evolution of the stellar
effective radius; \citet{KEB2017}, who explored the gas morphologies as a
function of galaxy mass; and \citet{ElBadry2018}, who showed that
$\mstar\sim10^8\msun$ galaxies are, on average, overly dispersion supported
relative to spatially unresolved HI gas kinematics.

However, more detailed comparison of our existing simulations to these observations 
is also warranted, particularly to forward-model the actual observed rotation curves 
and velocity dispersions. Some of the observed systems with high apparent velocities 
are clearly tidally disturbed or strongly interacting (e.g.\ IC 10, \citealt{Ashley2014}, 
and NGC~205, above), and \citet{Teyssier2012} argue NGC~147, 185, 6822, and Tucana, 
have all had a previous passage through the MW or M31 disk.  Some of these also feature 
recent starbursts, in which case \citet{ElBadry2017} argue that feedback-driven 
perturbations to the potential (the same which flatten the DM profile) can lead to the 
observationally-inferred Jeans masses (hence $\vcirc$) being over-estimated by up to 
a factor $\sim 2$ (sufficient to explain most of the discrepancy). We will show below, 
for example, that the actual line-of-sight stellar velocity dispersions in the 
simulations reach values similar to those observed even in the high-density systems.

\begin{figure*}
\centering
\includegraphics[width=0.33\textwidth]{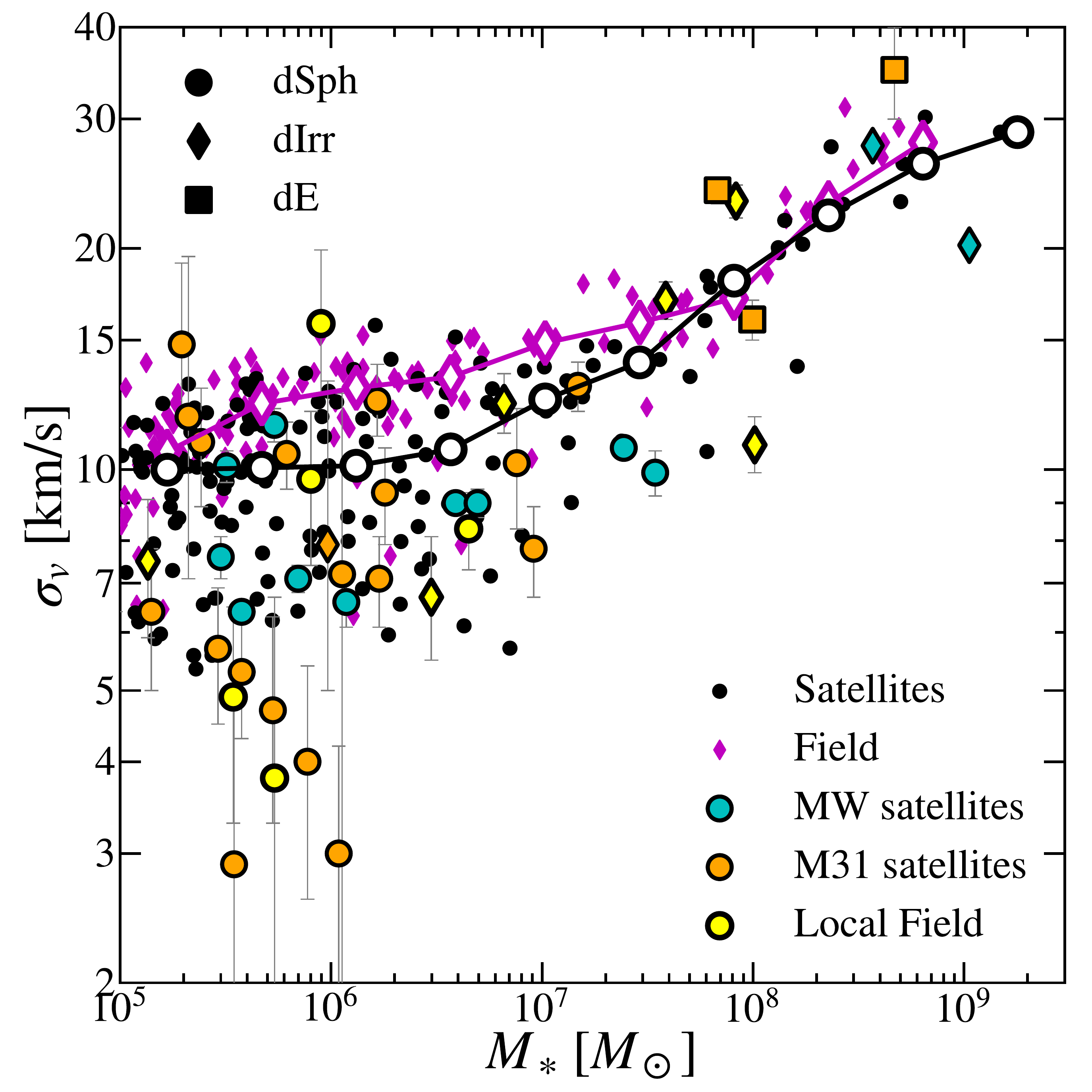}
\includegraphics[width=0.33\textwidth]{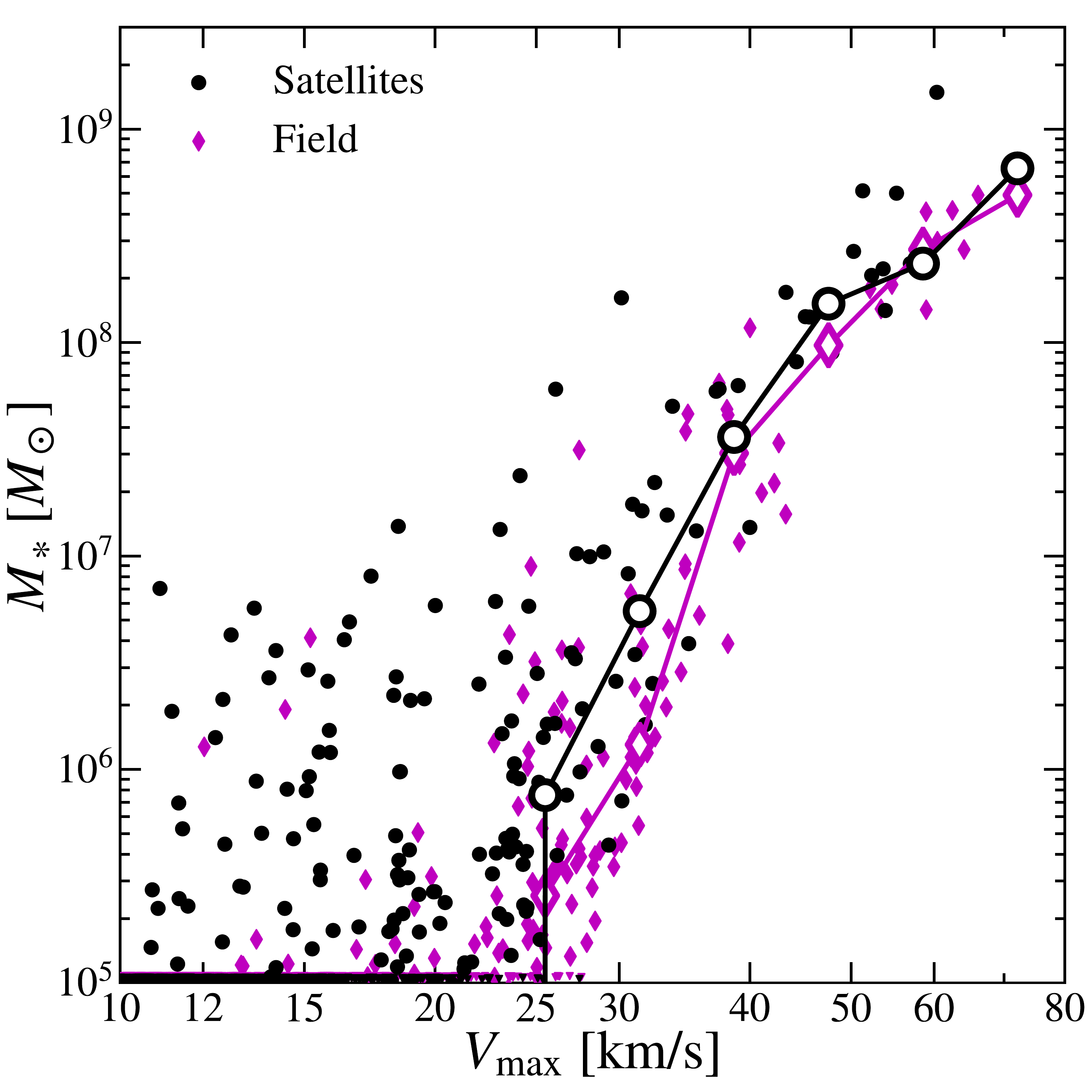}
\includegraphics[width=0.33\textwidth]{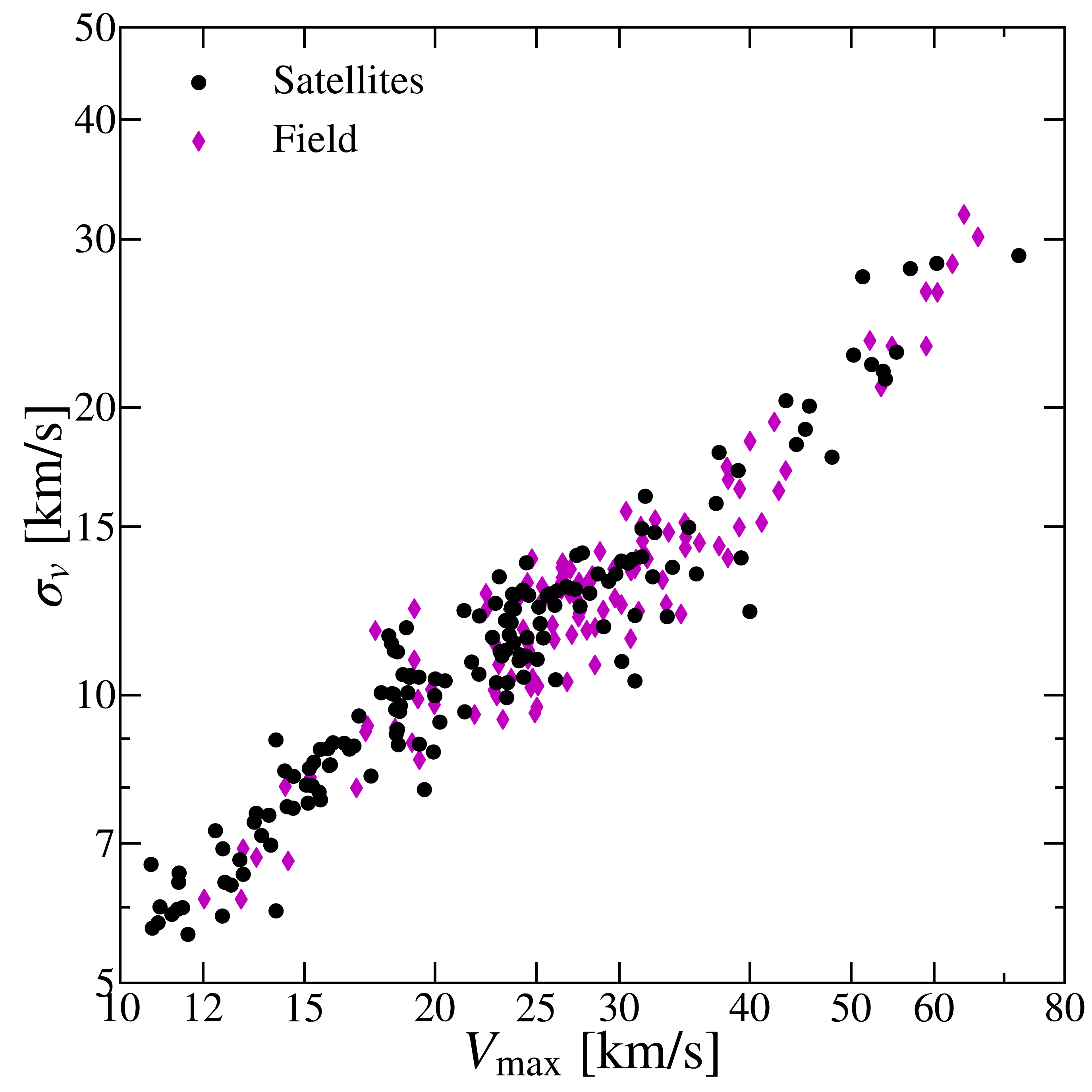}
\caption{
Relationships between stellar mass, 1D (line-of-sight) stellar velocity
dispersion, and halo $\vmax$ including satellites and Local Field galaxies from
all the simulations.  The simulations generally reproduce the observed
$\mstar-\sigstar$ relationship, particularly for the satellites of the MW, but
they fail to create any dwarfs with $\sigstar$ as low as some observed near M31,
possibly because of either artificial destruction (specifically, an inability to
track strongly tidally stripped objects) or N-body dynamical heating in the
simulations.  $\vmax$ is reasonably predictive of $\mstar$ for non-satellite 
galaxies, but tidal interactions decrease $\vmax$ faster than $\mstar$, generally 
scattering galaxies to the left.  $\vmax$ and $\sigstar$ remain remarkably 
correlated, however.  The open points in the left two panels 
indicate the medians for each population.  The downward arrows in the central 
panel indicate halos that fall off the plot (i.e. $\mstar<10^5\msun$), which 
first appear for $\vmax\lesssim25~\kms$ and become common at $\vmax\lesssim20~\kms$.  
We do not claim that these halos are necessarily ``dark,'' merely that they are 
at lower stellar mass.  For the purposes of calculating the medians in each 
population, these points are treated as having a stellar mass of zero.
}
\label{fig:scatters}
\end{figure*}

\subsection{The impact of the shape of the density profile}
\label{sssec:caveats} 
In summary, Figures~\ref{fig:vcirc_dmo}~and~\ref{fig:vcirc_hydro} demonstrate that, 
while the DMO analogues to the ELVIS on FIRE simulations all suffer from TBTF, the
problem is strongly alleviated or entirely eliminated in the fully hydrodynamic
runs.  Specifically, we find no  TBTF problem around the MW analogues, a result
consistent with  observational incompleteness in the Local Field, and a set of
dwarf galaxies consistent with the dSphs around M31 (though we find no analogues
to the higher density satellites of M31).

However, the analysis above was performed with two caveats: first, we assume NFW
profiles for the DMO (sub)halos but calculate $\vcirc$ for the FIRE simulations
by joining fitted density profiles to the raw particle data, and second, we
compare only the lower mass host in each pair to the MW dSphs.  The second
choice has no effect on our results:   by the metrics defined above, none of our
hosts, paired or isolated, have any massive failures in their luminous
satellites when compared with the MW dSphs.

The first choice is similarly irrelevant to our conclusions, but it does have
relatively large consequences for the number of `strong' massive failures
identified in the satellite populations of the DMO simulations:  without
correcting for the numerical impact of gravitational softening, we identify only
11 strong massive failures across the ten DMO hosts, compared with 61 when we
assume NFW profiles.  The number of massive failures in the DMO runs, however,
is much more stable to this assumption and only decreases by 1--4 in all  but
two of our hosts, with the total count decreasing by only $31\%$ from 114 to 79.
That is, by the \citet{ELVISTBTF} metrics, we would still have identified a TBTF
problem, even drawing directly from the particle data.  We also emphasize that
the assumption of NFW (or NFW-like) profiles for the DMO subhalos is
theoretically well-motivated.  Nevertheless, we plot the raw DMO $\vcirc$ curves
for \raj\ in Appendix~B for illustrative purposes.

More importantly, the results for the FIRE simulations are also only weakly
sensitive shape of the central density profile.  Specifically, adopting a cuspy
NFW profile \emph{vs.} using the corrected (or raw) mass profile has a
relatively minor influence on the number of massive failures identified in the
FIRE simulations, particularly when compared with the MW satellites.  Assuming
NFW profiles for the luminous satellites in the hydrodynamic runs (similarly
normalized to $\rmax$ and $\vmax$ of each subhalo) yields a total of only 13
massive failures across our ten hosts when compared with the MW dSph sample,
only three of which are `strong.'

Therefore, even though there is now substantial evidence that supernovae
feedback can flatten the central density profiles of $\mstar \simeq
10^{6.5}-10^9\msun$ galaxies \citep[e.g.][]{Pontzen2012, diCintio2014a,Chan2015}
we find that this effect is typically of second-order importance for solving
TBTF among the satellite populations in these simulations \citep[in agreement
with][]{Sawala2016}. Instead, the problem is primarily alleviated by removing
mass from the subhalos overall (lowering $\vmax$) and destroying otherwise
luminous satellites through enhanced tidal interactions with the disk
\citep{DOnghia2010,Sawala2016b,GKDisk}. However, we cannot completely dismiss
the importance of feedback induced core formation; for example, subhalos cored
by internal processes are then more susceptible to further mass loss from
external interactions (e.g. \citealp{Penarrubia2010,BrooksZolotov2012}, but also
see \citealp{GKDisk}, who showed that much of the differences in subhalo counts
between DMO and FIRE simulations can be accounted for purely by the
gravitational potential of the central galaxy with only a weak dependence on
subhalo mass or $\vmax$).

Changes to the internal profile are also relatively unimportant in the Local
Field, even though tidal effects are minimal in that volume:  assuming NFW
profiles for the non-satellite sample within $1~\mpc$ of each host increases the
total number of massive failures (defined in this volume as galaxies with
$\mstar\geq10^5\msun$ without observational kinematic counterparts) across the
entire simulated sample from $34$ to $40$. However, this difference is still
small compared to the overall impact of baryonic physics:  the same volumes
contain $\simeq75$ halos identified as massive failures when simulated without
baryons (nearly independent of whether we assume NFW profiles or use the raw
particle data).  Therefore, even in the Local Field, feedback induced cores are
only a small piece of resolving TBTF: overall baryonic mass loss, enhanced
disruption (both from other field galaxies and in the sample of  ``backsplash''
halos), and changes to the halo sample due to selecting on $\mstar$ rather than
$\vmax$ all play a significant role, even for non-satellite galaxies.

\section{Stellar velocity dispersions}
\label{ssec:scatters}

We have shown separately that the distributions of stellar masses and 
rotation curves of our simulated dwarf populations broadly agree with
that of the LG.  One can additionally ask whether our simulations
predict the correct joint relation between these; that is, whether our
individual dwarf galaxies are indeed realistic.
Figure~\ref{fig:scatters} directly compares the stellar
velocity dispersions (defined as the RMS line-of-sight dispersion of all the
stars associated with a galaxy) as a function of stellar mass for all of the
satellite galaxies (defined as $r<300~\kpc$) and non-satellite galaxies in the
simulations, together with dwarf galaxies from throughout the LG.  Though
$\sigstar$ is not necessarily representative of the underlying DM halo (e.g. in
the case of significant rotation, such as for the LMC, the right-most point in
the plot), the overall agreement between the simulated and observed
relationships support our assertion that our dwarf galaxy populations display
similar kinematics as the observed LG dwarf galaxies.

However, the simulations fail to reproduce the six LG galaxies with
$\mstar>3\times10^5\msun$ and $\sigstar\leq5~\kms$, all of which are within
$400~\kpc$ of M31.  This disagreement may indicate that our resolution (for the
stars, gravitational softening lengths $\lesssim5~\pc$ and particle masses
$\lesssim10^{3.5}$) remains insufficient for resolving the coldest, and
potentially most disrupted, dwarf galaxies in the LG -- these systems have
$\lesssim100$ star particles in the simulations.  The worst-case velocity kick
(i.e. the maximal possible deflection) due to N-body interactions between
stellar particles is of order $3~\kms$ in our simulations.  Therefore, it may
not be possible to maintain systems as dynamically cold as these six galaxies.
There is also evidence for a partial separation between the satellite and
non-satellite populations, such that satellite galaxies lose dynamical mass and
scatter to lower $\sigstar$ at fixed $\mstar$.  If, as suggested by
\citet{BrooksZolotov2012} and \citet{Zolotov2012}, this is due to tidal effects,
then the simulated analogues of the outlying galaxies in Figure~\ref{fig:scatters} 
may be (spuriously) destroyed due to finite mass resolution \citep{vandenBosch2018}.
However, those authors demonstrated artificial numerical disruption could
be minimized with aggressive gravitational force softenings, and we remind
the reader that our simulations adopt physical DM force softenings of 
$\simeq50~\pc$.

We also note that, while we do not plot it, our simulations typically agree
reasonably  well with the $\rhalf$ distribution of the LG population at fixed
$\mstar$ or $\sigstar$, but they do not reproduce the spatially smallest/most
compact systems at a given $\mstar$.  The results of even higher resolution FIRE
simulations of isolated dwarf galaxies suggest that our smallest simulated
dwarfs ($\mstar\lesssim10^6\msun$) will likely become more compact with
increased resolution \citep{Fitts2017}, but higher mass dwarf galaxies simulated
with FIRE maintain large effective radii even for gas particle masses
$260~\msun$ \citep{Chan2017} as their sizes are set by feedback ``puffing up''
the system. Given the insensitivity of our results to the internal profiles of
the simulated satellites, we do not expect that increasing the resolution will
significantly alter our conclusions with respect to TBTF.  Moreover, in lower 
resolution FIRE simulations, the higher mass (i.e. resolved) dwarf galaxies 
also yield a reasonable $\mstar-\sigstar$ relationship.

The right two panels in Figure~\ref{fig:scatters} plot the stellar mass and
stellar velocity dispersion as a function of $\vmax$.  The relationship between
$\mstar$ and $\vmax$ is relatively tight for isolated galaxies, where $\vmax$ is
more likely to represent the largest mass the halo ever reached, but it is clear
that tidal interactions shift galaxies to the left on the plot by removing dark
matter from the outer portions of the subhalos, decreasing $\vmax$ faster than
$\mstar$.\footnote{The relationship between $\mstar$ and $\vmax$ for
non-satellite galaxies is in stark contrast to the findings of
\citet{ELVISTBTF}, who found no trend between $\mstar$ and the implied $\vmax$
for galaxies in the Local Field. However, that analysis assumed fixed density
profiles across all halos, and assigned $\vmax$ by extrapolating from $\vhalf$.
An updated analysis that accounts for variance in the density profiles as a
function of $\mstar$ and $\mvir$ is required to properly assign $\vmax$ values
to the Local Field systems.} Both results are in good agreement with
\citet{Sawala2016}. Meanwhile, the relationship between $\vmax$ and $\sigstar$
remains remarkably tight even after tidal interactions with a larger halo.

The downward arrows at the bottom of the center panel indicate halos with
$\mstar<10^5\msun$, i.e. that fall below the $y$-limit of the plot.  Our
analysis assigns the vast majority of these halos no stars, though a few contain
a small number of star particles. These systems begin to appear for
$\vmax\lesssim25~\kms$ and become frequent for $\vmax\lesssim20~\kms$ (in rough
agreement with \citealp{Sawala2014}).  If these halos host ultra-faint dwarf
galaxies below our resolution limits, then such  galaxies should appear to be
fairly dense, with central masses similar to And XVIII.  Because our  definition
of ``massive failure'' includes only halos with $\vmax\geq25~\kms$, these dark
halos contribute only marginally towards resolving TBTF, particularly within the
virial radius of the MW.  However, the values plotted in
Figure~\ref{fig:scatters} are taken from the hydrodynamic simulations; it is
therefore possible that DMO halos with $\vmax\gtrsim25~\kms$ accreted less
overall mass in the hydrodynamic simulations and appear as dark halos with
$\vmax\lesssim20~\kms$.

\section{Conclusions}
\label{sec:conclusions}

The Local Group provides an unparalleled window into the population of dwarf
galaxies in the Universe, but it is not a typical environment:  the presence of
two massive halos ($\gtrsim10^{12}\msun$) has important implications for, e.g.,
the predicted halo mass function in the nearby volume \citep{ELVIS}.  Here, we
present the first two simulations from the ELVIS on FIRE suite, which apply the
FIRE models for star formation and feedback to LG-like environments at
$\lesssim4000\msun$ resolution.  We also include results from FIRE simulations
targeting isolated MW-mass halos at similar resolutions. We present the
satellite and non-satellite stellar mass functions predicted by these
simulations, and compare them to an analogous set of isolated MW-mass halos also
simulated with FIRE.  We then compare the internal structure of our resolved
galaxies to that of the dwarf galaxies in the LG, both via their implied
dynamical masses within the half-light radius (the too-big-to-fail problem) and
through the relationships between $\sigstar$ and $\mstar$.

The simulations accurately reproduce the dwarf galaxy population of the MW for
$\mstar\gtrsim10^5\msun$.  They roughly bracket the stellar mass function of
the MW satellites at nearly all masses, particularly below the masses of the
LMC and SMC.  However, the MW SMF is unique in exhibiting a ``gap'' between CVnI
($\mstar=3\times10^5\msun$) and the ultra-faint dwarfs with $\mstar \lesssim
3\times10^4\msun$, suggesting observational incompleteness around the MW even
for $\mstar>10^5\msun$ (typically $\approx4$ such galaxies).  The simulated
satellite galaxies also have central masses consistent with those of the real MW
satellites:  they do not suffer from too-big-to-fail.  This result is relatively
insensitive to the shape of the central density profile, particularly compared
to the total impact of baryonic physics:  even if we (falsely) assume a cuspy,
NFW profile for the hydrodynamic simulations, we identify less than two
``massive failures'' per host on average, while the DMO simulations contain more
than 11.  Therefore, supernova induced core formation is less important in resolving 
TBTF among the MW satellites: subhalo disruption and overall mass loss appear 
to be the dominant processes.

Our simulated satellites are somewhat less successful at reproducing the population 
of dwarf galaxies around M31.  They (usually) underproduce the total count at most 
stellar masses:  M31 contains, on average, roughly twice as many satellites with
$\mstar \geq 10^5\msun$ as the median simulated host.  Given that the highest mass
host in our sample has $\mvir=1.54\times10^{12}\msun$, this may suggest a higher
virial mass for M31.  Moreover, while our simulated satellites have central masses 
consistent with the dSphs around M31, none of our dwarf galaxies appears to have 
enough mass within $\sim300~\pc$ to host the highest-density dwarf galaxies inferred 
around M31 (the three dEs and IC 10) -- the opposite problem as TBTF. Our simulations
may also lack the resolution to reproduce the six dwarf galaxies within $400~\kpc$ of 
M31 with $\sigstar<5~\kms$ and $\mstar\geq3\times10^5\msun$.  More detailed modeling 
to predict the kinematics that would actually be measured in both these cases is 
clearly warranted.

The simulated non-satellite ($r_\mathrm{host}>300~\kpc$) populations agree
reasonably well with the observations:  they again roughly bracket the observed
SMFs, now for $\mstar\geq10^6\msun$, and have central masses that are consistent
with observations of the majority of the dwarf galaxies in the Local Field.
However, while the TBTF problem is resolved for satellite systems around the MW
and M31, the simulations predict the existence of $\sim10$ low-mass dwarf galaxies
within $\sim 1$Mpc of each host that are currently unaccounted for in the data.
These all have $\mstar = 10^5-10^6\msun$, and circular velocities broadly 
similar to those observed in other LG and Local Field dwarfs of the same mass, 
and thus may represent an as-of-yet undetected population of low-mass dwarf 
galaxies in the Local Field. This prediction should be testable with a combination 
of LSST, WFIRST, and thirty-meter class telescopes. However, we note that both this 
discrepancy and that with the M31 stellar mass function may be quantitatively 
reduced if some of our ``Local Field'' population should really be associated 
with M31 (in observations), and our non-paired halos demonstrate large systematic 
scatter in their field stellar mass functions.

Other than the very low $\sigstar$ dwarf galaxies near M31, our simulated dwarfs
broadly overlap the observations in $\mstar$ \emph{vs.} $\sigstar$.  We find a
tight relationship between $\sigstar$ and $\vmax$ for both satellites and
non-satellites.  The relationship between $\vmax$ and $\mstar$ is also
relatively tight for non-satellites, but tidal interactions introduce
substantial scatter among the satellite populations.

In short, neither the isolated, MW-mass FIRE simulations nor the ELVIS on FIRE
simulations suffer from the traditional small-scale problems identified for
satellites within the virial radius of the MW or M31.  Further, the ELVIS on
FIRE simulations alleviate the TBTF problem in the Local Field, though there
remains some tension that needs to be tested with future observations.  

Our simulations are not free of flaws.  They ignore some physical processes that
may be important at these scales (e.g. supermassive black holes and cosmic
rays), they include a reionization history that on the early edge of constraints
from Planck \citep{Onorbe2017}, they appear to lack the necessary resolution to
capture the half-mass radii of the smallest galaxies
($\mstar\lesssim10^6\msun$), and they may fail to reproduce the highest density
dwarf galaxies in the LG. Given that supernovae feedback appears to be most
effective at these mass scales, their existence may point towards physics
that reduces the burstiness in star formation (lessening the violent feedback 
episodes associated with strong bursts). Altogether, however, our results indicate
that a meta-galactic ionizing background, stellar/supernovae feedback, and
interactions with the disks of the MW and M31 are able to transform the overly
abundant, overly dense LG (sub)halo populations predicted by DMO simulations
into a sample of dwarf galaxies that is largely consistent with observations of
the LG within the vanilla $\Lambda$CDM paradigm, though our work does not rule
out non-standard DM physics.  Future work is required to fully understand the
relative contributions of internal feedback, LG-scale interactions, and the
cosmological background to dwarf galaxy formation in the Local Group, to test
the impact of host mass on the satellite populations, and to understand the
formation of the high density dEs in the Local Group.

\section*{Acknowledgments}

The authors thank Evan Kirby, Coral Wheeler, Lina Necib, Alejandro Benitez-Llambay, and Cameron Hummels for valuable discussions,
and Alexander Knebe and Oliver Hahn for making \texttt{AHF} and \texttt{MUSIC}, respectively, publicly available.

Support for SGK was provided by NASA through Einstein Postdoctoral Fellowship grant number PF5-160136 awarded by the Chandra X-ray Center, which is operated by the Smithsonian Astrophysical Observatory for NASA under contract NAS8-03060.
Support for PFH was provided by an Alfred P. Sloan Research Fellowship, NSF Collaborative Research Grant \#1715847 and CAREER grant \#1455342.
AW was supported by a Caltech-Carnegie Fellowship, in part through the Moore Center for Theoretical Cosmology and Physics at Caltech, and by NASA through grants HST-GO-14734 and HST-AR-15057 from STScI.
JSB was supported by NSF grant AST-1518291 and by NASA through HST theory grants (programs AR-13921, AR-13888, and AR-14282.001) awarded by STScI, which is operated by the Association of Universities for Research in Astronomy (AURA), Inc., under NASA contract NAS5-26555.
MBK acknowledges support from NSF grant AST-1517226 and CAREER grant AST-1752913 and from NASA grants NNX17AG29G and HST-AR-13888, HST-AR-13896, HST-AR-14282, HST-AR-14554, HST-AR-15006, HST-GO-12914, and HST-GO-14191 from STScI.
DK was supported by NSF grant AST-1715101 and the Cottrell Scholar Award from the Research Corporation for Science Advancement.
CAFG was supported by NSF through grants AST-1412836, AST-1517491, AST-1715216, and CAREER award AST-1652522, by NASA through grant NNX15AB22G, and by a Cottrell Scholar Award from the Research Corporation for Science Advancement.
KEB was supported by a Berkeley graduate fellowship, a Hellman award for graduate study, and an NSF Graduate Research Fellowship.
EQ was supported in part by NSF grant AST-1715070 and a Simons Investigator Award from the Simons Foundation.
RES is supported by an NSF Astronomy \& Astrophysics Postdoctoral Fellowship under grant AST-1400989.

Numerical calculations were run on the Caltech compute cluster ``Wheeler,''
allocations from XSEDE TG-AST130039 and PRAC NSF.1713353 supported by the
NSF, NASA HEC SMD-16-7223 and SMD-16-7592, and High Performance Computing
at Los Alamos National Labs.  This work also made use of \texttt{Astropy},
a community-developed core Python package for Astronomy \citep{Astropy},
\texttt{matplotlib} \citep{Matplotlib}, \texttt{numpy} \citep{numpy},
\texttt{scipy} \citep{scipy}, \texttt{ipython} \citep{ipython},
\texttt{yt} \citep{yt},  and NASA's Astrophysics Data System.  This research was supported in part by the National Science Foundation under Grant No. NSF PHY-1748958.

\bibliographystyle{mnras}
\bibliography{elvis_flagship}

\appendix

\section{OBSERVATIONAL CATALOG}
\label{sec:obsgals}

Table~\ref{tab:obs} lists the properties of the galaxies plotted in
Figures~\ref{fig:mstar},~\ref{fig:vcirc_dmo},~\ref{fig:vcirc_hydro},~and~\ref{fig:scatters},
separated in the MW, M31, and Local Field subsamples.  Columns list the distance
from the MW and M31, adopted stellar mass, 3D half-light radius, line-of-sight
velocity dispersion, the implied circular velocity at $\rhalf$, and references.
Galaxies without an entry for ($\rhalf$, $\vhalf$) are not included in
Figures~\ref{fig:vcirc_dmo}~and~\ref{fig:vcirc_hydro}, and galaxies without an
entry for $\sigstar$ are not included in Figure~\ref{fig:scatters}.  While we
include the properties of M32 as listed in \citet{Tollerud2014}, we remind the
reader that it falls outside the bounds of our plots in
Figures~\ref{fig:vcirc_dmo},~\ref{fig:vcirc_hydro},~and~\ref{fig:scatters}.
Dynamical values ($\rhalf$ and $\vhalf$) adopted from \citet{Wolf2010} were
calculated using data from \citet{Walker2009} along with \citet{Munoz2005,
Koch2007,Simon2007} and \citet{Mateo2008}.

The references in the last column are as follows:
(1) \citet{deVaucouleurs1991};
(2) \citet{Clementini2003};
(3) \citet{vanderMarel2002};
(4) \citet{Udalski1999};
(5) \citet{Harris2006};
(6) \citet{Monaco2004};
(7) \citet{Mateo1998};
(8) \citet{Frinchaboy2012};
(9) \citet{Pietrzynski2009};
(10) \citet{Wolf2010};
(11) \citet{Bellazzini2004};
(12) \citet{Pietrzynski2008};
(13) \citet{Bellazzini2005};
(14) \citet{Lee2009};
(15) \citet{Carrera2002};
(16) \citet{Bonanos2004};
(17) \citet{Martin2008CVnI};
(18) \citet{McConnachie2005};
(19) \citet{Simon2006};
(20) \citet{Geha2006};
(21) \citet{Fiorentino2010};
(22) \citet{Howley2012};
(23) \citet{Geha2010};
(24) \citet{Tikhonov2009};
(25) \citet{Wilcots1998};
(26) \citet{McConnachieIrwin2006};
(27) \citet{Tollerud2012};
(28) \citet{Martin2013};
(29) \citet{Conn2012};
(30) \citet{Ho2012};
(31) \citet{Richardson2011};
(32) \citet{Collins2013};
(33) \citet{Martin2009};
(34) \citet{Cook1999};
(35) \citet{Ibata2007};
(36) \citet{McConnachie2008};
(37) \citet{Brasseur2011};
(38) \citet{Bell2011};
(39) \citet{Tollerud2013};
(40) \citet{Collins2010};
(41) \citet{Yang2012};
(42) \citet{Chapman2013};
(43) \citet{Bernard2010};
(44) \citet{Kirby2014};
(45) \citet{Hunter2006};
(46) \citet{Gieren2006}
(47) \citet{Dale2007};
(48) \citet{Leaman2012};
(49) \citet{Bernard2009};
(50) \citet{Martin2013b};
(51) \citet{MartinezDelgado1999};
(52) \citet{Saviane1996};
(53) \citet{Fraternali2009};
(54) \citet{deJong2008};
(55) \citet{Simon2007}.

\begin{table*}
\begin{center}
\setlength\tabcolsep{6pt}
\def\arraystretch{1.025}%
\begin{tabular}[t]{llcccccc}
Name             & Type & $r_\mathrm{MW}$ & $r_\mathrm{M31}$ & $\mstar$           & $(\rhalf, \vhalf)$ & $\sigstar$ & References \\
                 &      & [$\kpc$]        & [$\kpc$]         & [$\msun$]          & [$\pc,\,\kms$]     & [$\kms$]   & \\
\hline\hline
\multicolumn{8}{c}{\textit{MW Satellites}}          \\
LMC              & dIrr & 50              & 811              & $1.1\times 10^{9}$ & --                 & 20.2       & 1, 2, 3 \\  
SMC              & dIrr & 61              & 812              & $3.7\times 10^{8}$ & --                 & 27.6       & 1, 4, 5 \\ 
Sagittarius      & dSph & 19              & 792              & $3.4\times 10^{7}$ & --                 & 9.9        & 6, 7, 8 \\  
Fornax           & dSph & 149             & 773              & $2.4\times 10^{7}$ & $(944,\,18.3)$     & 10.7       & 9, 10 \\ 
Leo I            & dSph & 257             & 922              & $4.9\times 10^{6}$ & $(388,\,15.7)$     & 9.0        & 10, 11 \\ 
Sculptor         & dSph & 86              & 766              & $3.9\times 10^{6}$ & $(375,\,16.1)$     & 9.0        & 10, 12 \\   
Leo II           & dSph & 236             & 902              & $1.2\times 10^{6}$ & $(233,\,11.6)$     & 6.6        & 10, 13  \\ 
Sextans          & dSph & 89              & 839              & $7\times 10^{5}$   & $(1019,\,12.1)$    & 7.1        & 10, 14  \\ 
Ursa Minor       & dSph & 78              & 758              & $5.4\times 10^{5}$ & $(588,\,20.2)$     & 11.5       & 10, 15 \\ 
Carina           & dSph & 107             & 842              & $3.8\times 10^{5}$ & $(334,\,11.1)$     & 6.4        & 9, 10  \\ 
Draco            & dSph & 76              & 755              & $3.2\times 10^{5}$ & $(291,\,17.7)$     & 10.1       & 10, 16 \\ 
CVnI             & dSph & 218             & 864              & $3\times 10^{5}$   & $(750,\,12.6)$     & 7.6        & 10, 17 \\ 

\multicolumn{8}{c}{\textit{M31 Satellites}}          \\
M33              & Spiral & 814             & 206              & $4.7\times 10^{9}$ & $(2344,\,50)$    & --         & 1, 18, 19 \\ 
NGC 205          & dE     & 828             & 42               & $4.7\times 10^{8}$ & $(520,\,41)$     & 35.0       & 1, 18, 20  \\ 
M32              & cE     & 809             & 23               & $4.1\times 10^{8}$ & $(110,\,79)$     & 92.0       & 1, 21, 22 \\ 
NGC 147          & dE     & 680             & 143              & $9.9\times 10^{7}$ & $(364,\,53)$     & 16.0       & 1, 18, 23 \\   
IC 10            & dIrr   & 798             & 252              & $7.7\times 10^{7}$ & $(612,\,35)$     & --         & 1, 24, 25 \\ 
NGC 185          & dE     & 621             & 188              & $6.8\times 10^{7}$ & $(295,\,52)$     & 24.0       & 1, 18, 23 \\ 
And VII          & dSph   & 765             & 218              & $1.5\times 10^{7}$ & $(972,\,23)$     & 13.0       & 18, 26, 27  \\  
And XXXII        & dSph   & 780             & 141              & $1.1\times 10^{7}$ & --                 & --         & 28 \\ 
And II           & dSph   & 656             & 184              & $9.1\times 10^{6}$ & $(1369,\,18.6)$    & 7.8        & 26, 29, 30 \\ 
And I            & dSph   & 749             & 58               & $7.6\times 10^{6}$ & $(839,\,18)$     & 10.2       & 18, 26, 27 \\ 
And XXXI         & dSph   & 760             & 263              & $6.5\times 10^{6}$ & --                 & --         & 28 \\ 
And III          & dSph   & 752             & 75               & $1.8\times 10^{6}$ & $(530,\,16)$     & 9.3        & 18, 26, 27 \\ 
And XXIII        & dSph   & 774             & 126              & $1.7\times 10^{6}$ & $(1335,\,12.3)$    & 7.1        & 29, 31, 32 \\  
And VI           & dSph   & 785             & 269              & $1.7\times 10^{6}$ & $(547,\,21.5)$     & 12.4       & 18, 26, 32 \\ 
And XXI          & dSph   & 831             & 134              & $1.1\times 10^{6}$ & $(1023,\,12)$    & 7.2        & 27, 29, 33 \\ 
And XXV          & dSph   & 817             & 89               & $1.1\times 10^{6}$ & $(853,\,5.2)$      & 3.0        & 29, 31, 32 \\  
LGS 3            & dE     & 773             & 269              & $9.6\times 10^{5}$ & $(626,\,9)$      & 7.9        & 18, 34 \\ 
And XV           & dSph   & 630             & 179              & $7.7\times 10^{5}$ & $(355,\,7)$      & 4.0        & 27, 29, 35 \\ 
And V            & dSph   & 777             & 109              & $6.2\times 10^{5}$ & $(442,\,18)$     & 10.5       & 18, 26, 27 \\ 
And XIX          & dSph   & 823             & 114              & $5.3\times 10^{5}$ & $(1972,\,8.1)$     & 4.7        & 29, 32, 36 \\ 
And XIV          & dSph   & 798             & 161              & $3.8\times 10^{5}$ & $(534,\,9)$      & 5.3        & 27, 29 \\ 
And XVII         & dSph   & 732             & 70               & $3.5\times 10^{5}$ & $(349,\,9.5)$      & 2.9        & 29, 32, 37 \\ 
And XXIX         & dSph   & 734             & 188              & $2.9\times 10^{5}$ & $(482,\,10)$     & 5.7        & 38, 39 \\ 
And IX           & dSph   & 770             & 40               & $2.4\times 10^{5}$ & $(726,\,19)$     & 10.9         & 18, 27 \\ 
And XXX          & dSph   & 686             & 148              & $2.1\times 10^{5}$ & $(356,\,20.6)$     & 11.8       & 29, 32 \\  
And XXVII        & dSph   & 832             & 74               & $2\times 10^{5}$   & $(875,\,25.3)$     & 14.8       & 29, 31, 32 \\ 
And XXIV         & dSph   & 605             & 208              & $1.5\times 10^{5}$ & --    & --         & 31 \\  
And X            & dSph   & 674             & 134              & $1.4\times 10^{5}$ & $(338,\,11.1)$     & 6.4        & 29, 32, 37 \\  
And XXVI         & dSph   & 766             & 103              & $9.6\times 10^{4}$ & $(296,\,14.9)$     & 8.6        & 32, 37 \\ 
And XI           & dSph   & 738             & 111              & $7.4\times 10^{4}$ & $(202,\,8)$      & 4.6          & 40, 41 \\ 
And XXII         & dSph   & 925             & 274              & $7.3\times 10^{4}$ & $(336,\,4.8)$      & 2.8        & 29, 33, 42 \\ 

\multicolumn{8}{c}{\textit{Local Field}}          \\
IC 1613          & dIrr & 758             & 520              & $10^{8}$   & $(1387,\,18.5)$    & 10.8       & 1, 43, 44, 45 \\ 
NGC 6822         & dIrr & 452             & 898              & $8.3\times 10^{7}$ & $(637,\,40.2)$     & 23.2       & 44, 45, 46, 47 \\ 
WLM              & dIrr & 933             & 836              & $3.9\times 10^{7}$ & $(2092,\,28.9)$    & 17.0       & 1, 18, 48 \\ 
Pegasus          & dIrr & 921             & 474              & $6.6\times 10^{6}$ & $(927,\,24.6)$     & 12.3       & 1, 18, 44, 45 \\ 
Cetus            & dSph & 756             & 680              & $4.5\times 10^{6}$ & $(816,\,14.5)$     & 8.3        & 26, 44, 49 \\  
Leo A            & dIrr & 803             & 1200             & $3\times 10^{6}$   & $(472,\,11.7)$     & 6.7        & 1, 44, 45 \\ 
And XXXIII       & dSph & 779             & 349              & $1.9\times 10^{6}$ & --                 & --         & 50 \\ 
Phoenix          & dIrr & 415             & 868              & $1.4\times 10^{6}$ & --                 & --         & 51 \\ 
Tucana           & dSph & 883             & 1356             & $9\times 10^{5}$   & $(279,\,33)$     & 15.8       & 49, 52, 53 \\ 
And XVIII        & dSph & 1217            & 453              & $8\times 10^{5}$   & $(417,\,17)$     & 9.7        & 27, 29, 36 \\  
And XVI          & dSph & 480             & 323              & $5.4\times 10^{5}$ & $(179,\,7)$      & 3.8        & 27, 29, 35 \\ 
And XXVIII       & dSph & 661             & 368              & $3.4\times 10^{5}$ & $(282,\,8)$      & 4.9        & 29, 35, 39 \\  
Leo T            & dIrr & 422             & 991              & $1.4\times 10^{5}$ & $(152,\,13)$     & 7.5        & 54, 55  \\ 
\end{tabular}
\end{center}
\caption{Observational properties of the galaxies included in our sample:
distance from the MW and M31, adopted stellar mass, position in $\vcirc(r)$
space, and line-of-sight stellar velocity dispersion.  References are listed in
Appendix~A.
}
\label{tab:obs}
\end{table*}

\section{SATELLITE GALAXIES WITHIN 400~KPC}
\label{sec:mstar400}

\begin{figure}
\includegraphics[width=\columnwidth,trim=0 3em 0 0]{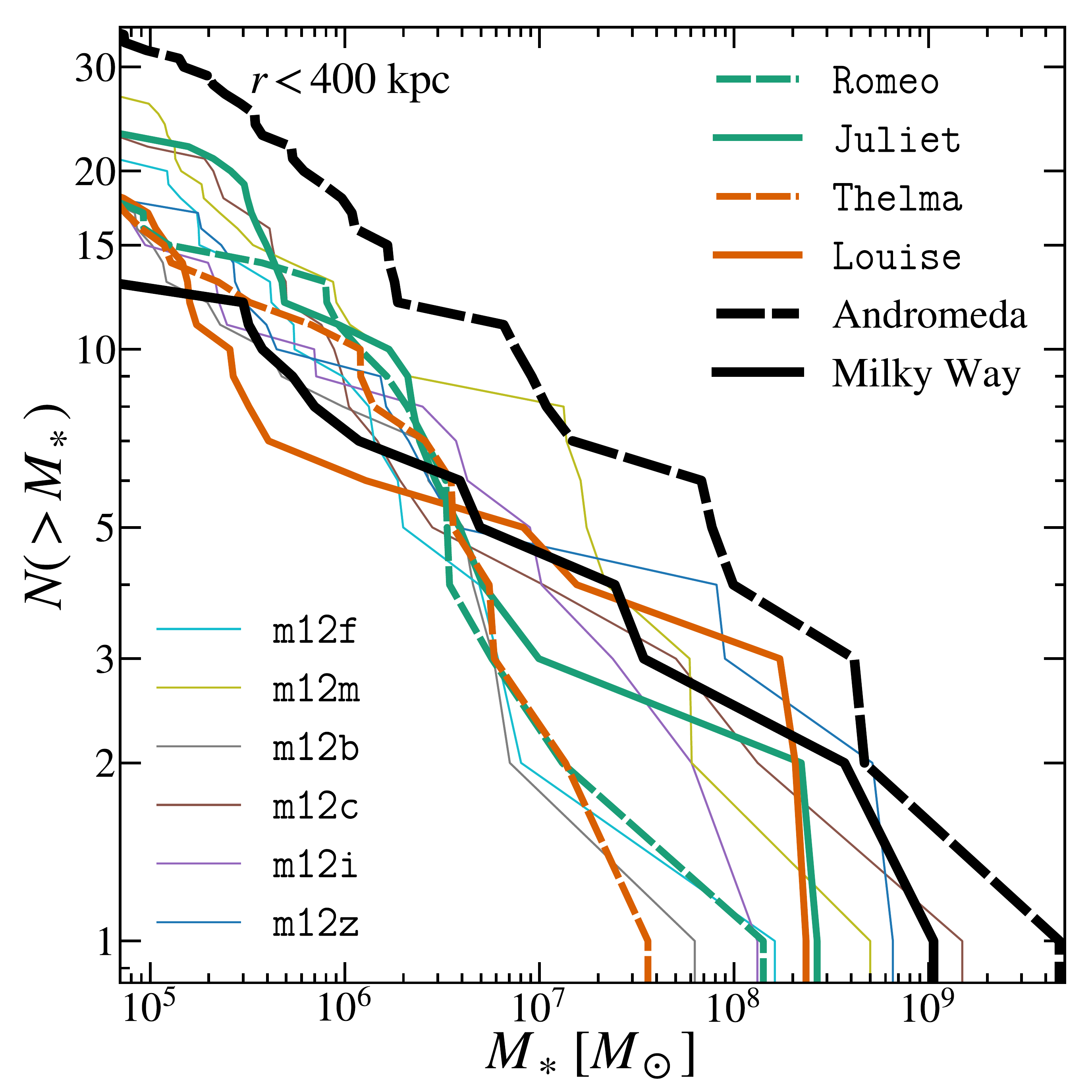}
\caption{Identical to the left panel of Figure~\ref{fig:mstar}, but 
counting galaxies within $400~\kpc$ (rather than $300~\kpc$).  Several of the
simulations (particularly \run{m12m}) are slightly closer to the SMF of M31 if
satellites are defined as $r<400~\kpc$, and this also reduces the tension with 
the stellar mass function (Figure~\ref{fig:mstar}, right panel) and TBTF 
comparison (Figure~\ref{fig:vcirc_hydro}, right panel) of the Local Field, 
by re-assigning a few halos to M31. The runs still tend to underpredict the 
SMF of M31, however.}
\label{fig:mstar400}
\end{figure}

In the main text, we select $r = 300~\kpc$ as the dividing radius between
satellites and non-satellite galaxies.  However, this is motivated primarily
by historical reasons, and is somewhat arbitrary~--~while the virial radius
of these hosts range from $\simeq240-300~\kpc$, the virial radius does not 
have an intrinsic physical meaning.  Instead, the physical boundary of the
halo is more closely related to the splashback radius, which is typically 
$\simeq0.8-1\times R_{200\mathrm{m}}$, the radius that encloses an average
of 200 times the background density \citep{More2015}.  For our hosts, 
$R_{200\mathrm{m}}\simeq310-380~\kpc$.  Moreover, there are curious gaps of 
known satellites at $257 - 415~\kpc$ around the MW and $274 - 349~\kpc$ 
around M31 (Samuel et al., in preparation).  Therefore, Figure~\ref{fig:mstar400} 
follows the left panel of Figure~\ref{fig:mstar} in counting galaxies around 
each host, but here counts dwarf galaxies within $400~\kpc$.  The exact 
numbers shift slightly, but the overall conclusion that our present sample 
of hosts (with $\mvir\leq1.45\times10^{12}\msun$) underproduces the SMF of 
M31 is unchanged. This does, however, somewhat reduce {\em both} this tension and 
the apparent tension with the Local Field stellar mass function and TBTF problems, by 
re-assigning $\sim5-10$ halos with $M_{\ast}\sim 10^{5}-10^{6}\,\msun$ from the Local Field to M31.

\section{IMPACT OF DENSITY PROFILES ON CIRCULAR VELOCITIES}
\label{sec:fitting}
\begin{figure*}
\includegraphics[width=\textwidth, trim=0 7em 0 0]{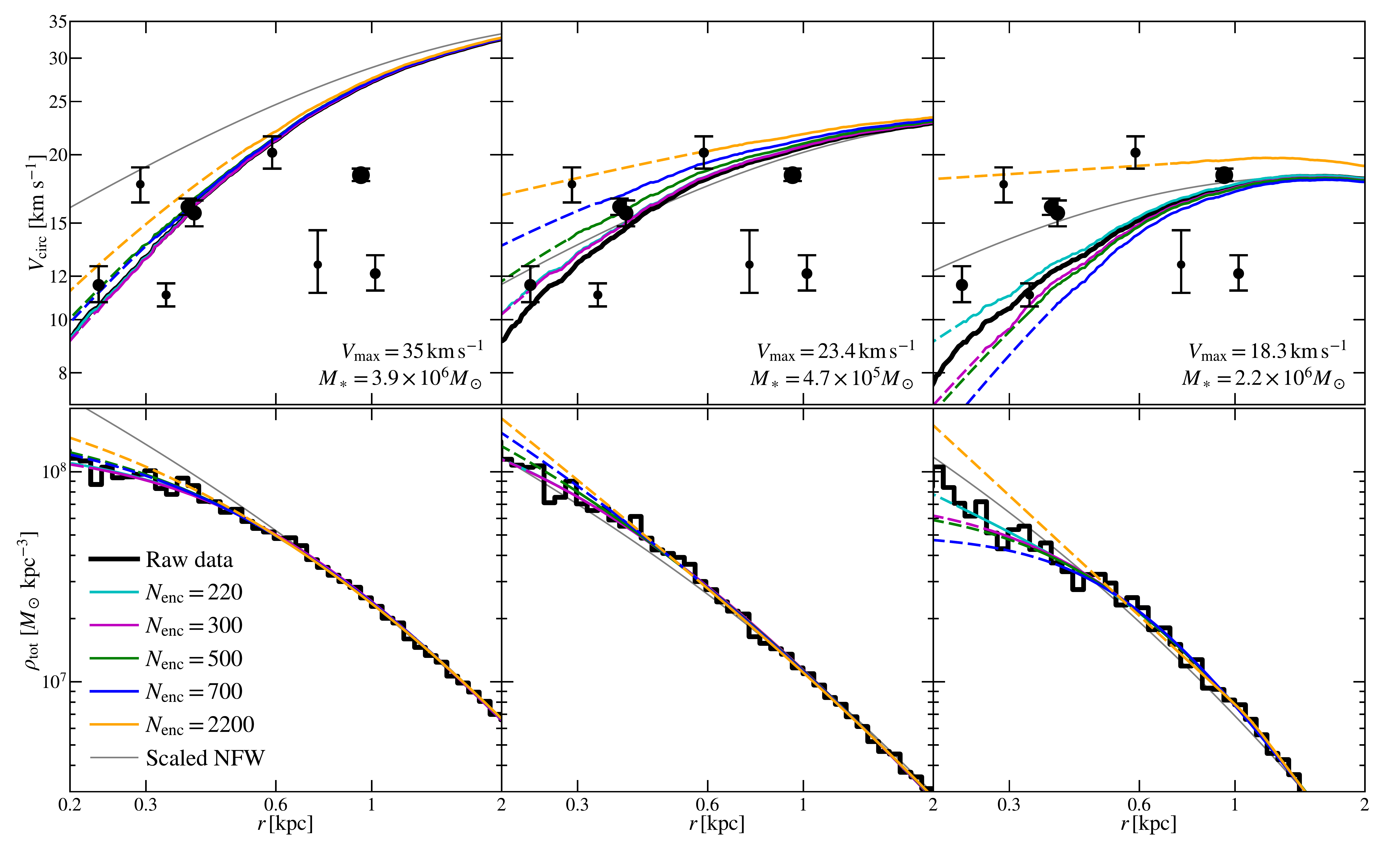}
\caption{Circular velocity curves (\emph{top}) and density profiles (\emph{bottom})
of three satellites around \run{Juliet} from the raw particle data, from
extrapolating fitted density profiles inside the radius enclosing
$N_\mathrm{enc}$ dark matter particles, and from assuming an NFW profile based
on $\rmax$ and $\vmax$, together with the usual constraints on the MW dSphs.
The colored lines become dashed at $r < r_\mathrm{min}$, i.e. where $\vcirc$ is
entirely determined by the fits.  The affects of varying $N_\mathrm{enc}$ are
negligible, particularly in the context of TBTF, provided that the implied
minimum radius is small enough to capture the curvature of the density profile.}
\label{fig:vcirc_comp}
\end{figure*}

\begin{figure*}
\includegraphics[width=\textwidth, trim=0 7em 0 0]{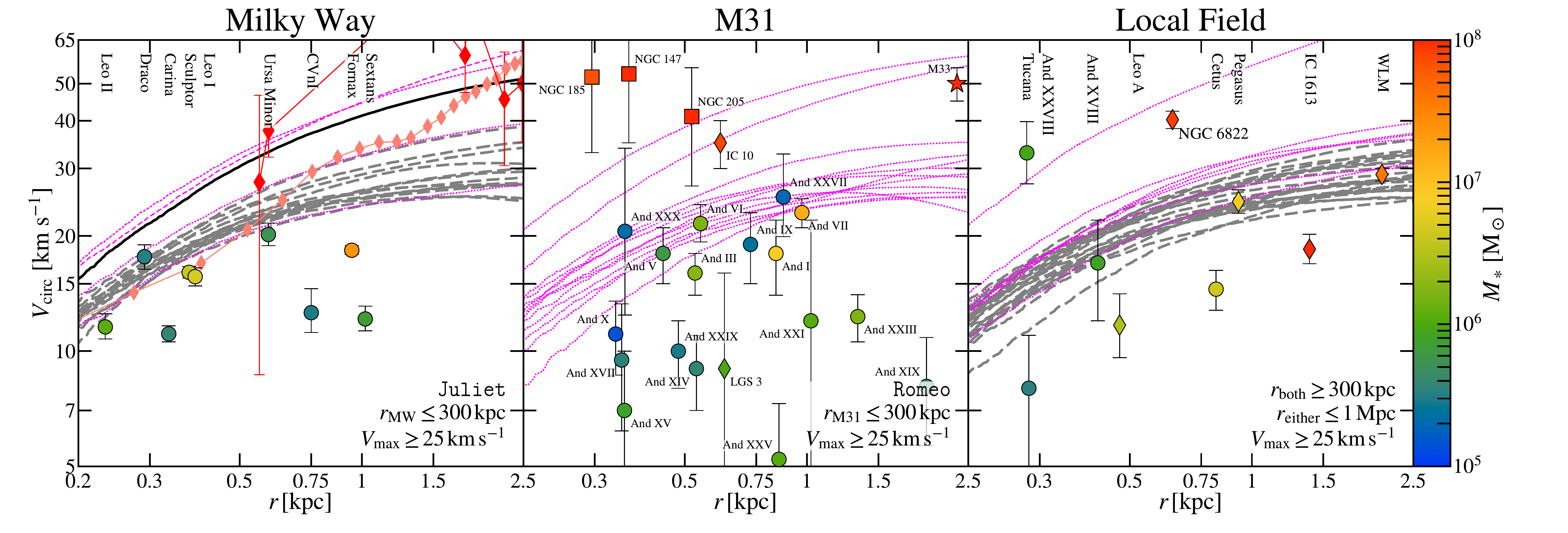}
\caption{Identical to the top panels of Figure~\ref{fig:vcirc_dmo}, but here using
the raw particle data for the DMO simulations.  The conclusions are unchanged:
the DMO simulations contain a wealth of subhalos with circular velocities at
$r\sim300-1500~\pc$ that are incompatible with the MW dSphs and the majority of
the M31 satellites.}
\label{fig:rawvcirc}
\label{lastpage}
\end{figure*}

In \S~\ref{ssec:tbtf}, we compute circular velocity profiles for galaxies in the
hydrodynamic simulations by first fitting the resolved portion of the halo with
an $(\alpha, \beta, \gamma)$ density profile:
\begin{equation}
\rho(r) = \rho_0 \left(\frac{r}{r_s}\right)^{-\gamma}\left[1 + \left(\frac{r}{r_s}\right)^\alpha\right]^{-(\beta-\gamma)/\alpha},
\end{equation}

Here, we examine the impact of using the interpolated density profile to compute
the central mass \emph{vs.} using the raw particle data, as well as the affect
of varying $r_\mathrm{min}$, the smallest radius used in fitting $\rho(r)$ and
the radius within which we compute $M(r)$ from the fits.
Figure~\ref{fig:vcirc_comp} shows the circular velocity profiles for three
satellite galaxies of \run{Juliet} obtained from the raw particle data and from
fits with varying $r_\mathrm{min}$. Following \citet[][and also see
\citealp{Power2003}]{FIRE2}, we determine $r_\mathrm{min}$ as the radius that
contains a given number of dark matter particles. Figure~\ref{fig:vcirc_comp}
illustrates that the correction to the circular velocity from under-resolving
the central $\sim100-500~\pc$ is typically only $\lesssim3~\kms$ at $500~\pc$
and is nearly negligible at $1~\kpc$, regardless of the number of DM particles
used to define $r_\mathrm{min}$.  The lone exception, the fit with
$r_\mathrm{min}$ determined by $N_\mathrm{enc}=2200$ in the right panel,
diverges because the implied $r_\mathrm{min}$ is comparable to $\rmax$,
resulting in a poor fit to the density profile.  To emphasize that our TBTF
counts are insensitive to the inner profiles, such that we can still match
constraints on the MW dSphs with cuspy central densities, we additionally plot
NFW profiles normalized to $\rmax$ and $\vmax$.  The implied $\vcirc$ profiles
can vary by a factor of $\sim2$ at $250~\pc$ in subhalos with higher $\mstar$
(where supernova feedback is more important), but a combination of overall mass
loss from surviving subhalos and enhanced subhalo destruction from the central
galaxy (relative to the DMO simulations) are nearly sufficient to explain TBTF.

In order to explicitly demonstrate that our assumption of NFW profiles for the
DMO simulations does not effect our overall results, Figure~\ref{fig:rawvcirc}
shows the circular velocity profiles for \raj, but here using the raw particle
data.  Even without correcting for the effects of gravitational softening or
assuming a density profile, the DMO simulations contain $\sim15$ subhalos that
can only host Draco and Ursa Minor.

\bsp	
\end{document}